\documentclass[fleqn,usenatbib]{mnras}

\usepackage[T1]{fontenc}

\DeclareRobustCommand{\VAN}[3]{#2}
\let\VANthebibliography\thebibliography
\def\thebibliography{\DeclareRobustCommand{\VAN}[3]{##3}\VANthebibliography}

\usepackage{graphicx}	
\usepackage{amsmath}	
\usepackage{amssymb}	
\usepackage{bm}

\usepackage{physics}
\renewcommand{\vec}{\bm}

\title{Inclination Dynamics of Resonant Planets under the Influence of an Inclined External Companion}

\author[Rodet \& Lai]{
Laetitia Rodet$^{1}$\thanks{E-mail: lbr63@cornell.edu},
Dong Lai$^{1,2}$
\\
$^{1}$Cornell Center for Astrophysics and Planetary Science, Department of Astronomy, Cornell University, Ithaca, NY 14853, USA\\
$^{2}$Tsung-Dao Lee Institute, Shanghai Jiao Tong University, Shanghai 200240, China
}

\date{Accepted XXX. Received YYY; in original form ZZZ}

\pubyear{2020}

\begin{document}
\label{firstpage}
\pagerange{\pageref{firstpage}--\pageref{lastpage}}
\maketitle

\begin{abstract}
	 Recent observations suggest that a large fraction of Kepler super-Earth systems have external giant planet companions (cold Jupiters), which can shape the architecture of the inner planets, in particular their mutual inclinations. The dynamical perturbation from cold Jupiters may account for the population of misaligned planets in the Kepler data. The effectiveness of this mechanism can be hindered by a strong planet--planet coupling in the inner system. In this paper, we study how mean-motion resonances (MMRs) affect this coupling and the expected misalignment. We derive approximate analytical expressions for the mutual inclination excitations in the inner planet system induced by an inclined companion, for various period ratios and perturber properties. In most cases, the mutual inclination is proportional to a dimensionless parameter that characterizes the strength of the perturber relative to the coupling in the inner system. We show that the MMR strengthens the inner coupling, reducing the mutual inclination induced by the perturber by a factor of a few. We find that the resonance is resilient to the perturbation, and derive a criterion for the libration of the resonant angle. Our results have applications for constraining unseen planetary perturbers, and for understanding the architecture of multiplanet systems.
\end{abstract}




\section{Introduction}

\label{sec:intro}

Among the thousands of exoplanets detected to date, the most common are super-Earths or mini-Neptunes \citep{2011ApJS..197....8L,2014ApJ...790..146F,2015ARA&A..53..409W}. They have masses and sizes between that of the Earth and Neptune, and orbital periods below 300 days. About 30 \% of Sun-like stars host super-Earths, with each system containing an average of three planets \citep{2018ApJ...860..101Z}. These planetary systems are generally “dynamically cold”, with eccentricities $e \sim 0.02$ and mutual inclinations $I \sim 2^\circ$ \citep[e.g.,][]{2015ARA&A..53..409W}. However, there is an observed excess of single transiting planets, that could be sign of a dynamically hot sub-population of misaligned planets \citep[so-called Kepler dichotomy,][]{2011ApJS..197....8L,2012ApJ...758...39J,2017MNRAS.469..171R,2019MNRAS.490.4575H}. There is also an indication that the mutual inclination dispersion in a given system increases rapidly with decreasing number of planets \citep{2018ApJ...860..101Z}. In addition, a fraction of super-Earth systems have orbital periods close to mean-motion resonances \citep[MMRs; see e.g.][]{2015ARA&A..53..409W,2020MNRAS.495.4192C}.

In recent years, long-period giant planets (i.e. “Cold Jupiters” or CJs) have been observed in an increasing number of super-Earth systems. Statistical analyses combining radial velocity and transit observations confirmed a strong correlation between the occurrence of super-Earth systems and CJs: depending on the metallicities of their host stars, 30--60 \% of inner super-Earth systems have CJ companions  \citep{2018AJ....156...92Z, 2019AJ....157...52B}. Massive companions may shape the architecture of super-Earth systems through various dynamical processes. It has been suggested that stars with CJs or with high metallicities have smaller multiplicity of inner super-Earths \citep{2018AJ....156...92Z}. Moreover, \cite{2020AJ....159...38M} found  that  these  CJ  companions  are  typically  misaligned with their inner systems (mutual inclination of $I_p \sim 12$ deg on average).
 
The dynamical perturbations from external companions can excite the eccentricities and mutual inclinations of super-Earths, thereby influencing the observability (co-transiting geometry) and stability of the inner system. This subject has been studied extensively in recent years, in a general framework \citep[e.g.;][]{2014ApJ...789..111B, 2016MNRAS.463.3226C, 2017AJ....153...42L, 2017AJ....153..210H, 2017MNRAS.468.3000M, 2017MNRAS.467.1531H, 2017MNRAS.468..549B, 2017MNRAS.469..171R, 2018MNRAS.478..197P, 2019MNRAS.482.4146D, 2020arXiv200805698P} or in connection to specific systems with misalignment and obliquity measurements or detected cold companions \citep[e.g.;][]{2017AJ....153..227J,2017MNRAS.464.1709G,2020AJ....159..242W,2020MNRAS.497.2096X,2020arXiv200706410D,2020MNRAS.496.4688P}. Kepler-88 for example comprises two inner planets, misaligned by $3^\circ$, and an outer giant companion \citep{2020AJ....159..242W}.

In this paper, we will consider two inner super-Earths perturbed by a misaligned CJ. The theoretical framework has been studied previously in \cite{2014ApJ...789..111B} and \cite{2017AJ....153...42L} in particular. They found that the outer body induces a differential nodal precession in the inner system, which introduces a misalignment between initially coplanar super-Earths. Nevertheless, the mutual inclination between the inner planets can remain small if their mutual coupling is strong enough to temper the perturber's influence.

However, these analyses focused on pure secular interactions, and are no longer valid when the inner planets have commensurable periods, i.e. when they are in MMR. While the analysis of Kepler's period ratio distribution does not evidence a strong excess of commensurability, the proximity to MMR has been found in systems with external CJs, such as the Kepler-88 system mentioned above, where the two inner planets are close to the $2:1$ MMR. While MMRs seem resilient to transient perturbations such as moderately close encounters \citep[i.e. not enough to trigger ejections; see ][]{2020MNRAS.496.1149L}, it is unclear to what extent an external companion can affect the inner MMRs of super-Earths.

In this paper, we carry out a theoretical study to determine how the mutual inclination of two initially coplanar inner planets in MMR is influenced by an inclined external perturber. As a related issue, we will determine the effect of the perturber on the resonance. We focus on inclination dynamics, and restrict the analysis to circular orbits. Eccentricity will be taken into account in a future paper. Order two MMRs are the strongest for which the resonance is coupled to the inclination. We will focus on this type of resonances in this paper.

Section \ref{sec:setup} describes the setup and characteristic timescales (frequencies) of the problem, and the anticipated results. In Section \ref{sec:testmass-planet}, we consider the restricted 3-body problem, where a test mass is in MMR with an inner planet, perturbed by an external body. We present the equations of motion derived from the Hamiltonian, the approximate analytical solution and a comparison with the numerical results. We show that the MMR increases the coupling between the inner planets, and thus temper the ``disruptive'' influence of the external perturber. We also derive a condition for the resonance to hold (i.e. the resonant angle to librate) throughout the precession cycle. Section \ref{sec:testmass-planet_inner} presents the case where the test mass is inner to the planet. In this case, a secular resonant feature can trigger strong misalignments for ``intermediate'' couplings. In Section \ref{sec:planet-planet}, we generalize these results to the problem of two finite-mass planets in MMR. We summarize our key findings and conclude in Section \ref{sec:conclusion}.

\section{Set-up, Timescales and Coupling Parameters}
\label{sec:setup}

Consider two planets (masses $m_1$ and $m_2$) in circular orbits (semi-major axes $a_1$ and $a_2$) around a central star (mass $M$). The orbital angular momenta are denoted by $\vec{L_1} = L_1 \vec{\hat l_1}$ and $\vec{L_2} = L_2 \vec{\hat l_2}$, where $\vec{\hat l_1}$ and $\vec{\hat l_2}$ are unit vectors. The two planets are initially close to coplanarity. An external perturber (mass $m_{\rm p}$) moves in a circular inclined orbit, with semi-major axis $a_{\rm p}$, orbital angular momentum $\vec{L_{\rm p}} = L_{\rm p} \vec{\hat{l}_{\rm p}}$, and inclination $I_{\rm p}$ with respect to the inner planets' initial orbital plane. Throughout the paper, we assume $L_{\rm p} \gg L_1, L_2$, so that $\vec{L_{\rm p}}$ is constant in time. The question we are interested in is: How does the mutual inclination $I$ of the two inner planets evolve?

Let us define the dimensionless masses and Keplerian orbital frequencies (mean-motions) of the planets:
\begin{equation}
\mu_k = \frac{m_k}{M},\quad n_k = \sqrt{\frac{GM}{a_k^{3}}},
\end{equation}
where $k=1,2,{\rm p}$ labels the different bodies.

The inner planets revolve around the central star following a nearly Keplerian motion, perturbed by both their fellow inner planet and the outer companion. The dominant inclination evolution is associated with the secular precession of the longitude of the node, i.e. the slow rotation of the planet's orbital plane. Denote $\omega_{ik}$ the characteristic precession frequency of planet $i$ induced by the gravitational torque from planet $k$. The frequencies of mutual interactions in the inner two planets are
\begin{align}
\omega_{12} & \equiv  \frac{1}{4}n_2\mu_2\sqrt{\frac{a_1}{a_2}} \,b_{\frac{3}{2}}^{(1)}\!\!\left(\frac{a_1}{a_2}\right), \\
\omega_{21} & \equiv \frac{\omega_{12} L_1}{L_2} =   \frac{1}{4}n_1\mu_1 \left(\frac{a_1}{a_2}\right)^{\frac{5}{2}}  b_{\frac{3}{2}}^{(1)}\!\!\left(\frac{a_1}{a_2}\right),
\end{align}
while the precession frequencies induced by the perturber are 
\begin{align}
\omega_{1{\rm p}} &\equiv \frac{1}{4}n_{\rm p}\mu_{\rm p} \sqrt{\frac{a_1}{a_{\rm p}}}\, b_{\frac{3}{2}}^{(1)}\!\!\left(\frac{a_1}{a_{\rm p}}\right),\\
\omega_{2{\rm {\rm p}}} &\equiv \frac{1}{4}n_{\rm p}\mu_{\rm p} \sqrt{\frac{a_2}{a_{\rm p}}} \,b_{\frac{3}{2}}^{(1)}\!\!\left(\frac{a_2}{a_{\rm p}}\right),
\end{align}
\citep[see Eq.~7.11 in][]{2000ssd..book.....M}. The $b_s^{(j)}$ functions are the Laplace coefficients:
\begin{equation}
b_s^{(j)}(\alpha) = \frac{1}{\upi}\int_{0}^{2\upi} \frac{\cos(jx)\mathrm{d}x}{(1-2\alpha\cos x +\alpha^2)^s}.
\end{equation}

\cite{2017AJ....153...42L} considered pure secular interactions between the planets and the perturber. They showed that the inclination dynamics of the inner planets depend on the dimensionless coupling parameter
\begin{equation}
\varepsilon_{12} = \frac{\omega_{2{\rm p}}-\omega_{1{\rm p}}}{\omega_{21}+\omega_{12}},\label{eq:eps12}
\end{equation}
which measures the relative strength of the perturber with respect to the coupling in the inner system. For $\varepsilon_{12} \ll 1$, the inner planets are strongly coupled and remain ``rigidly'' coplanar; for $\varepsilon_{12}\gg 1$, the planets are weakly coupled, and each undergoes free nodal precession driven by the perturber.

In this paper, we consider the situation where the two planets are near the $j:(j-2)$ MMR, i.e. $(j-2)n_1 \approx jn_2$. This MMR introduces two resonant coupling frequencies between the two planets, given by \citep[see Eqs.~8.19 and 8.20 in][]{2000ssd..book.....M} 
\begin{align}
\omega_{12,\mathrm{res}} &\equiv \frac{1}{4}n_2\mu_2 \sqrt{\frac{a_1}{a_2}}\, b_\frac{3}{2}^{(j-1)}\!\!\left(\frac{a_1}{a_2}\right), \\
\omega_{21,\mathrm{res}} &\equiv  \frac{\omega_{12, \mathrm{res}} L_1}{L_2} = \frac{1}{4}n_1\mu_1\left(\frac{a_1}{a_2}\right)^\frac{5}{2}  b_\frac{3}{2}^{(j-1)}\!\!\left(\frac{a_1}{a_2}\right).
\end{align}
Note that these resonant coupling frequencies have a similar order of magnitude as the secular frequencies:
\begin{align}
\frac{\omega_{12, \mathrm{res}}}{\omega_{12}} &= \frac{\omega_{21, \mathrm{res}}}{\omega_{21}} = \frac{b_{\frac{3}{2}}^{(j-1)}\!(\frac{a_1}{a_2})}{b_{\frac{3}{2}}^{(1)}\!(\frac{a_1}{a_2})} \approx 0.5
\end{align}

\begin{figure}
	\centering
	\includegraphics[width=0.9\linewidth]{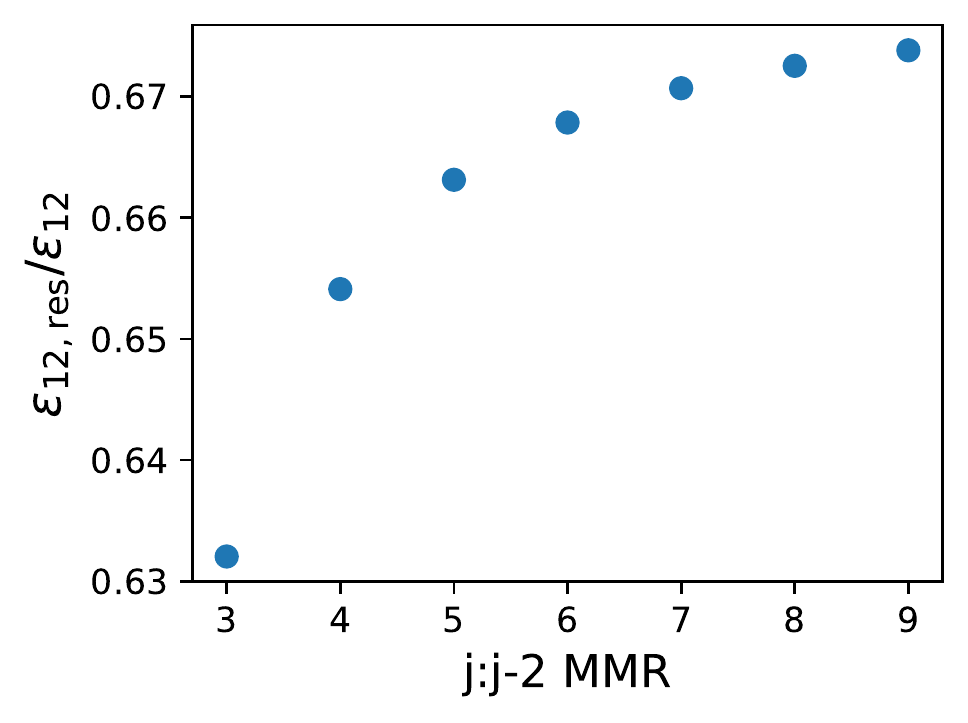}
	\caption{Perturber's strength ratio $\varepsilon_{12\mathrm{res}}/\varepsilon_{12}$ (Eq.~\ref{eq:eps12res}) as a function of the resonance parameter $j$ (corresponding to the $j:j-2$ MMR). }\label{fig:eps12res}
\end{figure}

In this new study, we will show that in the presence of MMR, the dimensionless coupling parameter becomes
\begin{equation}
\varepsilon_{12, \mathrm{res}} = \frac{\omega_{2{\rm p}}-\omega_{1{\rm p}}}{\omega_{21} + \omega_{21,\mathrm{res}}+\omega_{12} +\omega_{12,\mathrm{res}}} \approx 0.65~ \varepsilon_{12}\label{eq:eps12res}.
\end{equation}
The ratio between $\varepsilon_{12, \mathrm{res}}$ and $\varepsilon_{12}$ depends only slightly on $j$ (see Fig.~\ref{fig:eps12res}).
The difference from the non-resonant case (Eq.~\ref{eq:eps12}) is the resonant-induced precession terms $\omega_{12,\mathrm{res}}$ and $\omega_{21,\mathrm{res}}$ in the denominator. As these terms are positive, the MMR enhances the coupling between planets and further tempers the influence of the outer perturber. For $a_{\rm p} \gg a_2$, $\varepsilon_{12,\mathrm{res}}$ is proportional to $\mu_{\rm p}/a_{\rm p}^3$. For the $1:3$ MMR, it is approximately given by:
\begin{align}
\varepsilon_{12,\mathrm{res}} &\approx 0.1 \left(\frac{\mu_p}{100\mu}\right) \left(\frac{10 a_2}{a_{\rm p}}\right)^3,
\end{align}
where $\mu$ is the largest of $\mu_1$ and $\mu_2$.

\section{Test mass-planet resonance perturbed by an external body: outer test mass} \label{sec:testmass-planet}

In order to have a theoretical understanding of the problem, we will first simplify it by supposing that one of the planets has negligible mass. Here we present the case where planet $2$ is a test mass ($m_2 \ll m_1, m_{\rm p}$). For convenience, we adopt the notations that the primed quantities represent the finite-mass planet and the unprimed quantities the test mass: $(a_1=a') < (a_2=a) < a_{\rm p}$. 

\begin{figure}
	\includegraphics[width=\linewidth]{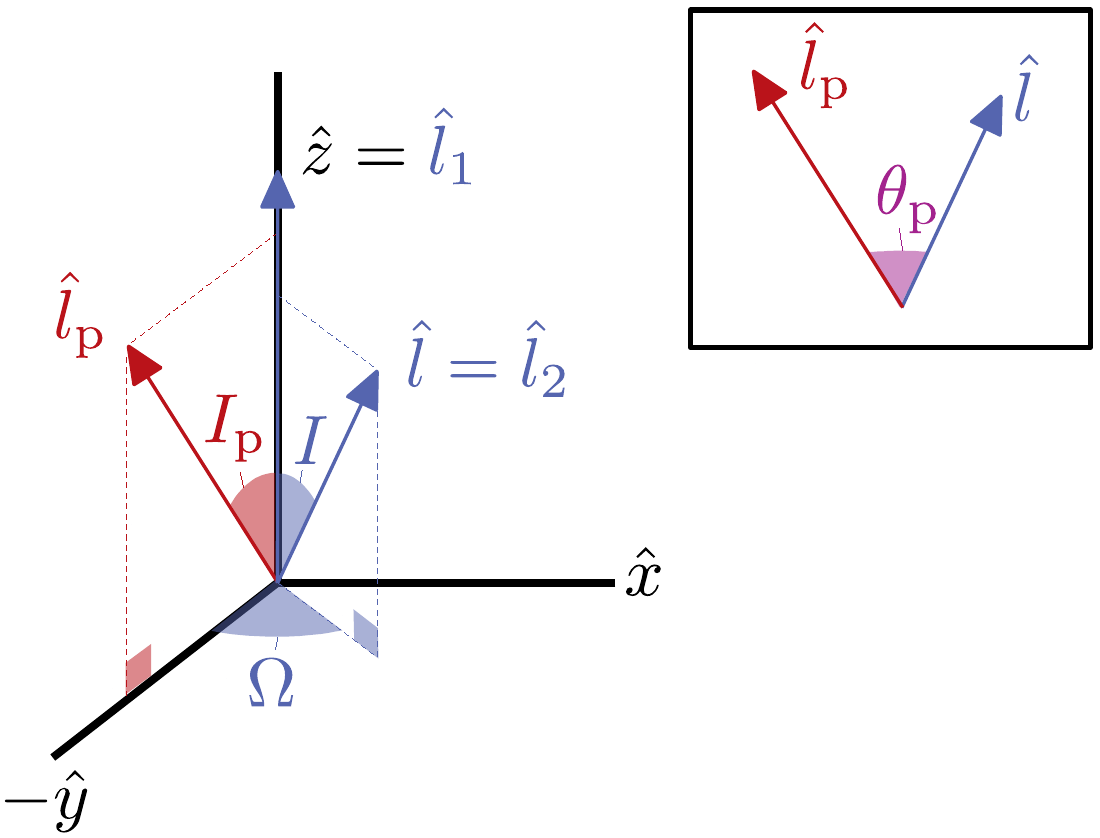}
	\caption{Coordinate system used in Section \ref{sec:testmass-planet}. The z-axis is along $\vec{\hat l_1}$, the orbital angular momentum unit vector of $m_1$ (the finite-mass planet), and $\vec{\hat l_{\rm p}}$ and $\vec{\hat l}$ specify the orbital angular momentum axes of $m_{\rm p}$ (the external perturber) and $m$ (the test mass). This frame is not inertial: it is rotating uniformly due to the influence of $m_{\rm p}$.}\label{fig:notationsvectors}
\end{figure}

Let us study the problem in the orbital plane of planet $1$, which is rotating uniformly due to the influence of the perturber. We position the frame such that the ascending node of the outer perturber corresponds to the origin of longitude ($\Omega_{\rm p} = 0$, see Fig.~\ref{fig:notationsvectors}).

\subsection{Hamiltonian and evolution equations}

In the purely secular problem, the inclination dynamics is governed by the evolution of the angular momentum axis of each planet, so that a vector approach is convenient \citep{2017AJ....153...42L}. However, near MMR, the system has an additional degree of freedom, the resonant angle, that is not accounted for in the vector approach. In the following, we use Hamiltonian mechanics, with the expression of the Hamiltonian valid for $I \ll 1$. 

Let us write down the Hamiltonian per unit mass of the test mass in the orbital plane of the inner finite-mass planet, using the canonical variables
\begin{align}
\lambda&,~  \Lambda = \sqrt{GMa},\\
-\Omega&,~  Z =\sqrt{GMa}(1-\cos I) \simeq \sqrt{GMa}{I^2}/{2},
\end{align}
where $\lambda$ is the mean longitude and $\Omega$ the longitude of the ascending node. The Hamiltonian has several pieces. First, at order 2 in $I$, the interactions between the inner planets and with the central star give \citep[Eq.~8.20 in][]{2000ssd..book.....M}:
\begin{equation}
H_{21} = -\frac{GM}{2a} - \frac{1}{2}\omega_\mathrm{res} \Lambda I^2\cos\phi + \frac{1}{2}  \omega_{21} \Lambda I^2,
\end{equation}
where the second and third terms represent respectively the resonant ($\omega_\mathrm{res} \equiv \omega_\mathrm{21,res}$) and secular interactions on the test mass ($m_2$) from the planet ($m_1$). The resonant angle $\phi$ is defined as
\begin{align}
\phi = j\lambda_2 - (j-2)\lambda_1 - 2\Omega.
\end{align}
Second, in the presence of the outer perturber $m_{\rm p}$, the secular interaction on $m$ from $m_{\rm p}$ gives 
\begin{equation}
H_{2{\rm p}} = \frac{1}{2} \omega_{2{\rm p}} \Lambda \sin^2 \theta_{\rm p},
\end{equation}
where $\theta_{\rm p}$ is the angle between the orbital planes of $m$ and $m_{\rm p}$, which is given by
\begin{align}
\cos\theta_{\rm p} = \cos I \cos I_{\rm p} + \cos \Omega \sin I \sin I_{\rm p}.\label{eq:thetap}
\end{align}
Finally, the frame that we are in (Fig.~\ref{fig:notationsvectors}), where $\vec{L_1}$ and $\vec{L_{\rm p}}$ are constant, rotates at the frequency $-\omega_{1{\rm p}}\cos I_{\rm p}$ around $\vec{\hat l_{\rm p}}$, so that a term must be added to the Hamiltonian \citep{2014AmJPh..82..769T}:
\begin{equation}
H_\mathrm{rot} = (\omega_{1{\rm p}}\cos I_{\rm p}) \vec{\hat l_{\rm p}}\cdot\frac{\vec{L_2}}{m} =  \omega_{1{\rm p}}\Lambda\cos I_{\rm p} \cos\theta_{\rm p}.
\end{equation}
Note that an additional secular interaction term $H_\mathrm{sec, 0}(a)$ that does not depend on any angle should be theoretically accounted for. However, it plays little role in the dynamics, only shifting the resonant location by a negligible amount. We neglect it in our analysis.

To proceed, it is convenient to make a canonical transformation to a new set of variables, replacing the fast varying $\lambda$ for the more slowly varying resonant angle $\phi$. The new angle-action pairs are
\begin{align}
\theta_1 = \phi &,\quad J_1 = \Lambda/j,\\
\theta_2 = \Omega&,\quad J_2 = \Lambda(2/j - I^2/2),
\end{align}
obtained via the generating function
\begin{equation}
S = \left[j\lambda - (j-2)\lambda' - 2\Omega\right]J_1 + \Omega J_2.
\end{equation}
This adds a new term to the Hamiltonian, to take into account the time dependence of $\lambda' = n't$:
\begin{equation}
H_S = \pdv{S}{t} = -\frac{j-2}{j}n'\Lambda.
\end{equation}
Thus, the Hamiltonian (in the rotating frame) for the test mass due to the interaction with $M$, $m'=m_1$, and $m_{\rm p}$ is
\begin{equation}
H = H_{21} + H_{2{\rm p}} + H_\mathrm{rot} + H_S.
\end{equation}

The Hamilton-Jacobi equations to the first orders in $I$, $\mu'$ and $\mu_p$ give
\begin{align}
\dv{\phi}{t} &= \pdv{H}{J_1}\big|_{J_2} = j\pdv{H}{\Lambda} + j\frac{J_2}{I \Lambda^2}\pdv{H}{I},\label{eq:dphidtfromJ1}\\
\dv{\Omega}{t} &= \pdv{H}{J_2}\big|_{J_1} = -\frac{1}{I\Lambda} \pdv{H}{I},\\
\dv{J_1}{t} &= -\pdv{H}{\phi}\big|_\Omega = -\frac{1}{2}\omega_{\mathrm{res}}\Lambda I^2 \sin\phi,\\
\dv{J_2}{t} &= -\pdv{H}{\Omega}\big|_\phi = \Lambda I \sin I_{\rm p} \sin\Omega \left(\omega_{2{\rm p}} \cos\theta_{\rm p} - \omega_{1{\rm p}}  \cos I_{\rm p} \right).
\end{align}
On the other hand, we can retrieve the evolution of $a$ and $I$ from the evolution of $J_1$ and $J_2$:
\begin{align}
\dv{a}{t} &=  \frac{2j \Lambda}{GM} \dv{J_1}{t},\\
\dv{I}{t} &= -\frac{1}{\Lambda I} \dv{J_2}{t}.\label{eq:dIdtfromJ2}
\end{align}
Using Eqs.~\eqref{eq:dphidtfromJ1}--\eqref{eq:dIdtfromJ2}, to the first orders in $I$, we obtain a set of differential equations:
\begin{align}
\dv{a}{t} &= -j \omega_{\mathrm{res}} a I^2 \sin\phi\label{eq:dadt},\\
\dv{I}{t} &=  -\omega_{\mathrm{res}} I \sin\phi + \frac{\Delta\omega_{\rm p}}{2} \sin(2I_{\rm p}) \sin\Omega \label{eq:dIdt},\\
\dv{\phi}{t} &= -\Delta n - 2\omega_{\mathrm{res}} \cos\phi- \frac{\Delta\omega_{\rm p}}{I}\sin(2I_{\rm p})\cos\Omega\label{eq:dphidt},\\
\dv{\Omega}{t} &= \omega_{\mathrm{res}}\cos\phi -\omega_{21} - \Delta\omega_{\rm p}\cos ^2 I_{\rm p} + \frac{\Delta\omega_{\rm p}}{2I}\sin(2I_{\rm p})\cos\Omega\label{eq:dOdt},
\end{align}
where $\Delta\omega_{\rm p} = \omega_{2{\rm p}}-\omega_{1{\rm p}}$ is the relative precession frequency induced by the outer perturber, and $\Delta n$ characterizes the distance from the nominal resonant location:
\begin{equation}
\Delta n \equiv (j-2)n_1 - j n_2. \label{eq:eta}
\end{equation}

In the absence of external perturber, $\Delta\omega_{\rm p} = 0$, and the Hamiltonian does not depend on $\Omega$. Thus, the conjugated momentum $J_2$ is constant, which is equivalent at first orders to the conservation of the quantity
\begin{equation}
(a_1/a_2)(1+jI^2/2) \equiv \alpha_0 . \label{eq:alpha0}
\end{equation}
In that case, Eqs.~\eqref{eq:dadt} and \eqref{eq:dIdt} are not independent. As the evolution of $\Omega$ does not impact the overall dynamics, we are then left with a ``classic'' one-degree of freedom problem. However, in the presence of the perturber, $\alpha_0$ depends strongly on $\Omega$, and all equations (\ref{eq:dadt}--\ref{eq:dOdt}) must be taken into account.

The full system of differential equations is not solvable analytically without further approximation: the precession and resonance are strongly entangled. Of course, the equations can be integrated numerically.

\subsection{Decoupling resonance and precession}

\begin{figure}
	\centering
	\includegraphics[width=\linewidth]{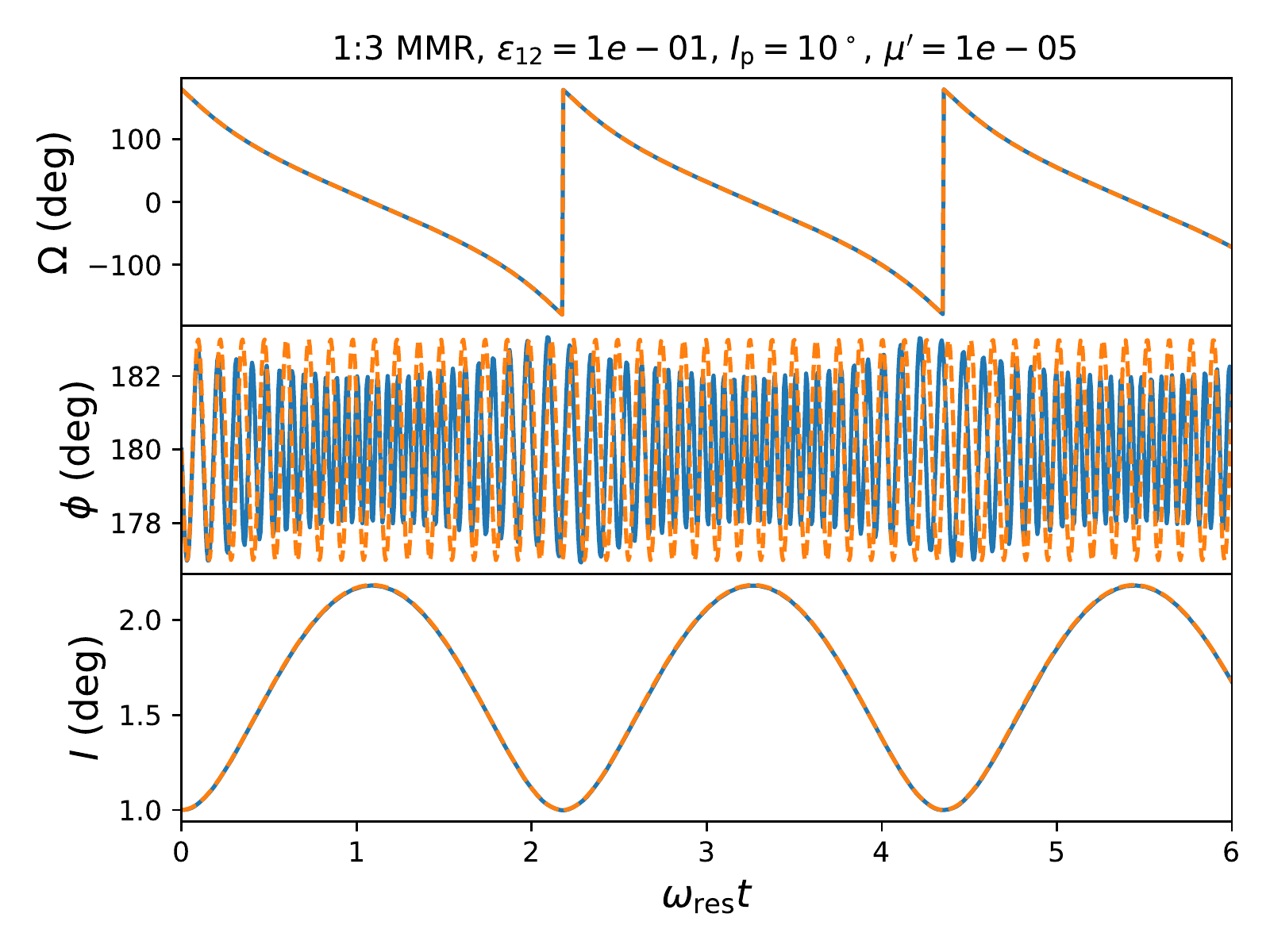}
	\caption{Typical evolution of a system with the inner pair of planets in 1:3 MMR, where $m_2$ is a test mass and initially inclined at $I(t=0) = 1^\circ$. The external perturber is inclined at $I_{\rm p} = 10^\circ$. This configuration has $\varepsilon_\mathrm{12} = 0.1$ (corresponding to $\varepsilon_\mathrm{12,res} \approx 6.10^{-2}$) and $\mu' = 10^{-5}$   (corresponding to $\omega_\mathrm{res} \approx 10^{-6}n'$). The blue lines correspond to the result of the numerical integration of Eqs.~\eqref{eq:dadt}--\eqref{eq:dOdt}, while the orange dotted lines represent the analytical approximation (Eq.~\ref{eq:secular} has been used for $\Omega$ and $I$, and the solution of Eq.~\ref{eq:d2thetadt2} for $\phi$, neglecting the varying terms that involve $\Omega$). The resonant angle varies much faster than the precession: the two motions are decoupled. Throughout the integration, the stability condition (\ref{eq:stability}) is satisfied by a great margin, so that the resonant angle oscillates around the equilibrium.} \label{fig:example}
\end{figure}

Our interest here is to determine the amplitude and frequency of the inclination variations, due to both the resonance and the precession. Equations (\ref{eq:dadt}--\ref{eq:dOdt}) can be solved numerically. A typical evolution is shown in Fig.~\ref{fig:example}. The semi-major axis oscillates very close to the resonance location, and is not shown here.

\begin{figure}
	\centering
	\includegraphics[width=\linewidth]{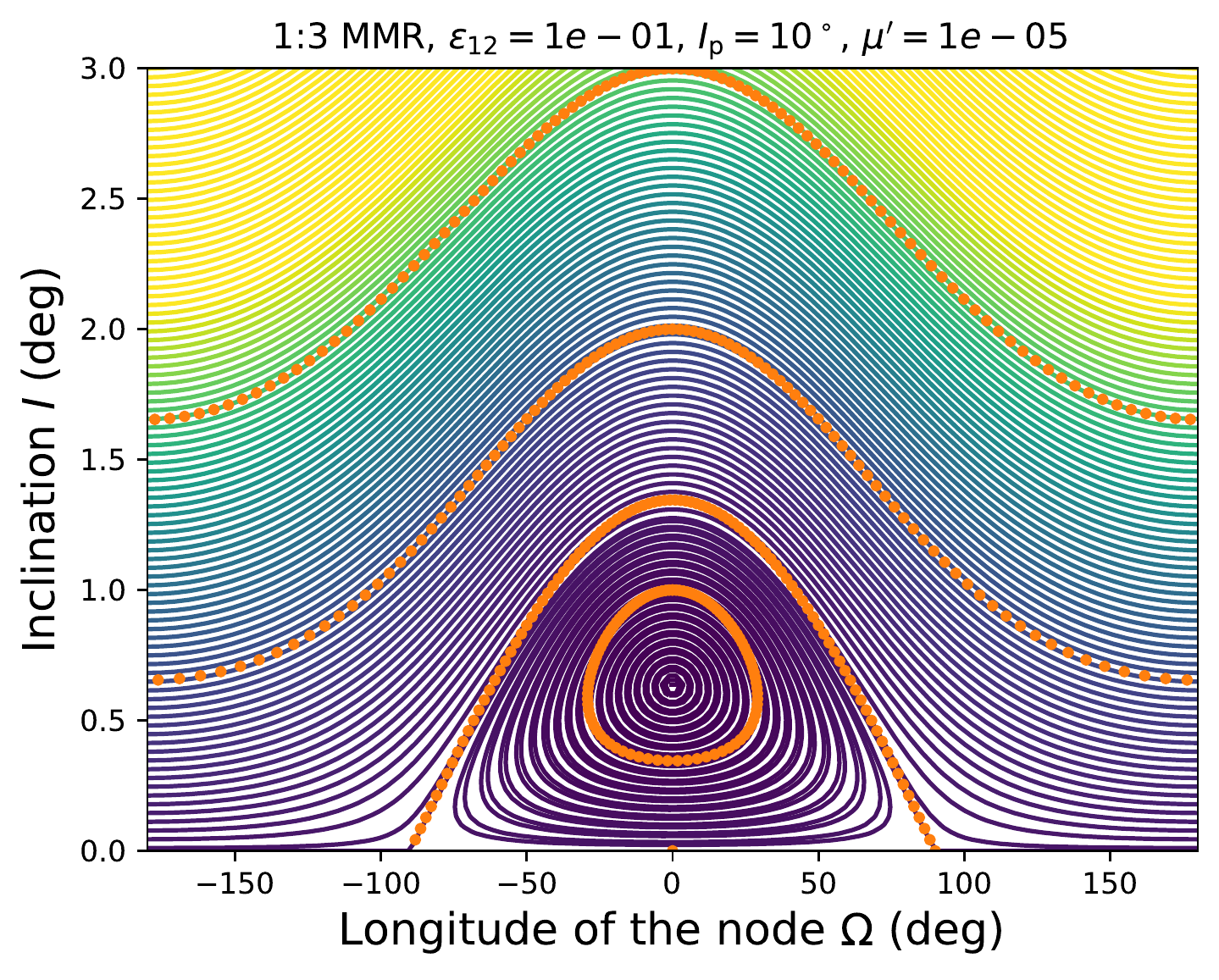}
	\caption{Hamiltonian map in the $\Omega-I$ plane, assuming $\phi = \upi$, with the same parameters as in Fig.~\ref{fig:example}. The orange dotted line represents the analytic solution Eq.~\eqref{eq:secular} with different initial conditions.}\label{fig:hmap}
\end{figure}

From Eq.~\eqref{eq:dOdt}, we can introduce the ``total'' precession frequency $\omega_{\mathrm{prec}}$ of the test mass in the rotating frame:
\begin{align}
\omega_{\mathrm{prec}} &\equiv  \omega_{\mathrm{res}} +\omega_{21} +\left(\omega_{2{\rm p}}-\omega_{1{\rm p}}\right)\cos^2 I_{\rm p}. \label{eq:wprec}
\end{align}

Now, let us study the decoupled problem. We assume that the resonance is at the equilibrium $\phi = \upi$ (we will give conditions on the validity of this hypothesis below). The evolution of the system on the secular timescale is then governed by:
\begin{align}
\dv{a}{t} &= 0,\\
\dv{I}{t} &=  \frac{\Delta\omega_{\rm p}}{2} \sin(2I_{\rm p})\sin\Omega\label{eq:didtsecular},\\
\dv{\Omega}{t} &= -\omega_{\mathrm{prec}} + \frac{\Delta\omega_{\rm p}}{2I}\sin(2I_{\rm p})\cos\Omega. \label{eq:dOdtsecular}
\end{align}
This problem is identical to the well-known forced eccentricity problem. The equations can be rewritten in the complex form:
\begin{equation}
\dv{}{t} \left(I e^{i\Omega}\right) = -i\omega_{\mathrm{prec}}I e^{i\Omega} + i \frac{1}{2}\Delta\omega_{\rm p}\sin(2I_{\rm p}).
\end{equation}
As $\Delta\omega_{\rm p}\sin(2I_{\rm p})$ and $\omega_{\mathrm{prec}}$ are constant, this differential equation is solvable, giving:
\begin{equation}
I e^{i\Omega} = I_{\mathrm{free}} e^{i(-\omega_{\mathrm{prec}}t + \psi)} +  I_{\mathrm{forced}} \label{eq:secular},
\end{equation}
where
\begin{align}
I_{\mathrm{forced}} &= \frac{1}{2} \frac{\Delta\omega_{\rm p}}{\omega_{\mathrm{prec}}}\sin(2I_{\rm p}), \label{eq:iforced}\\
&= 
\frac{1}{2}\varepsilon_{12,\mathrm{res}} \sin(2I_{\rm p})~\text{if}~\varepsilon_{12}\ll 1 .
\end{align}
is the forced inclination, with $\varepsilon_\mathrm{12,res}$ given by Eq.~\eqref{eq:eps12res}. The constant $I_\mathrm{free}$ (``free inclination'') and the phase $\psi$ are determined by the initial conditions.

The secular behaviour is thus a libration or circulation of the longitude of the node coupled to an oscillation of the inclination around the forced inclination. When the longitude of the node circulates, i.e. when $I_\mathrm{free} > I_\mathrm{forced}$, the maximum inclination amplitude is independent of the initial conditions and is given by
\begin{align}
I_\mathrm{max}-I_\mathrm{min} = 2I_\mathrm{forced}.\label{eq:imax}
\end{align}

Figure \ref{fig:hmap} represents the Hamiltonian map of the problem for a given $\varepsilon_{12}$. The analytical solution (Eq.\ref{eq:secular}) is also shown for several different initial conditions. Figure \ref{fig:example} depicts the agreement between the exact numerical solution and the analytical solution for the decoupled problem.

\begin{figure}
	\centering
	\includegraphics[width=\linewidth]{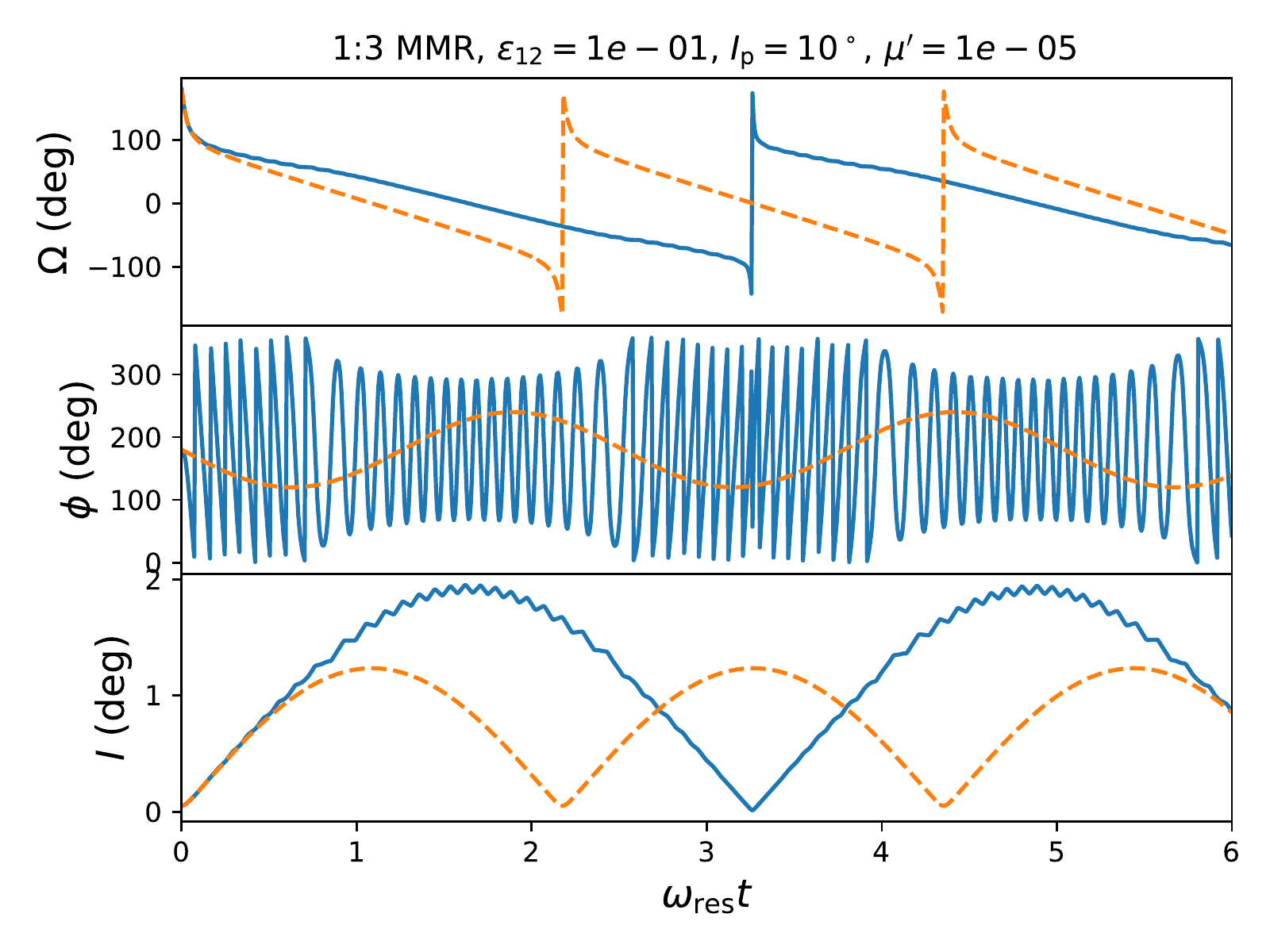}
	\caption{Same as Fig.~\ref{fig:example} except for the initial condition $I(t=0) = 0.05^\circ$. In this example, the condition (\ref{eq:stability}) is not ensured throughout the integration, so that the MMR is broken (the resonant angle circulates) and the solution for the inclination (orange lines, Eq.~\ref{eq:secular}) does not hold.}\label{fig:example_unstable}
\end{figure}

\begin{figure}
	\centering
	\includegraphics[width=\linewidth]{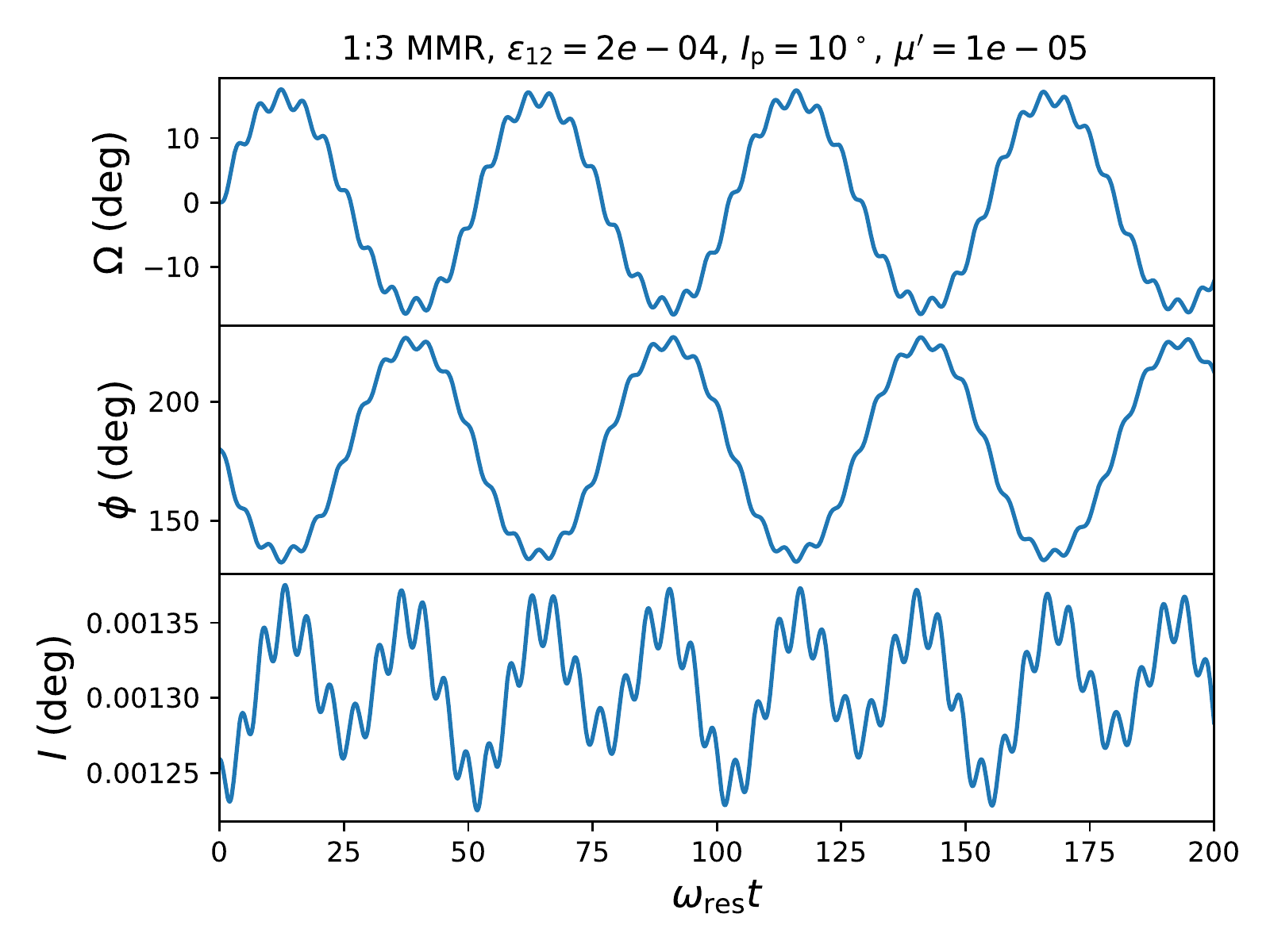}
	\caption{Same as Fig.~\ref{fig:example} except for $\varepsilon_\mathrm{12,res} \approx 10^{-4}$ and $I(t=0) = I_\mathrm{forced} \approx 0.001^{\circ}$ (Eq.~\ref{eq:iforced}). In this case, the resonant angle and the precession have similar timescales. The stability condition (\ref{eq:stability}) is not satisfied, but the stability is predicted by the more rigourous criterion (\ref{eq:instability}).} \label{fig:example_stabilityisland}
\end{figure}

\begin{figure*}
	\centering
	\includegraphics[width=\linewidth]{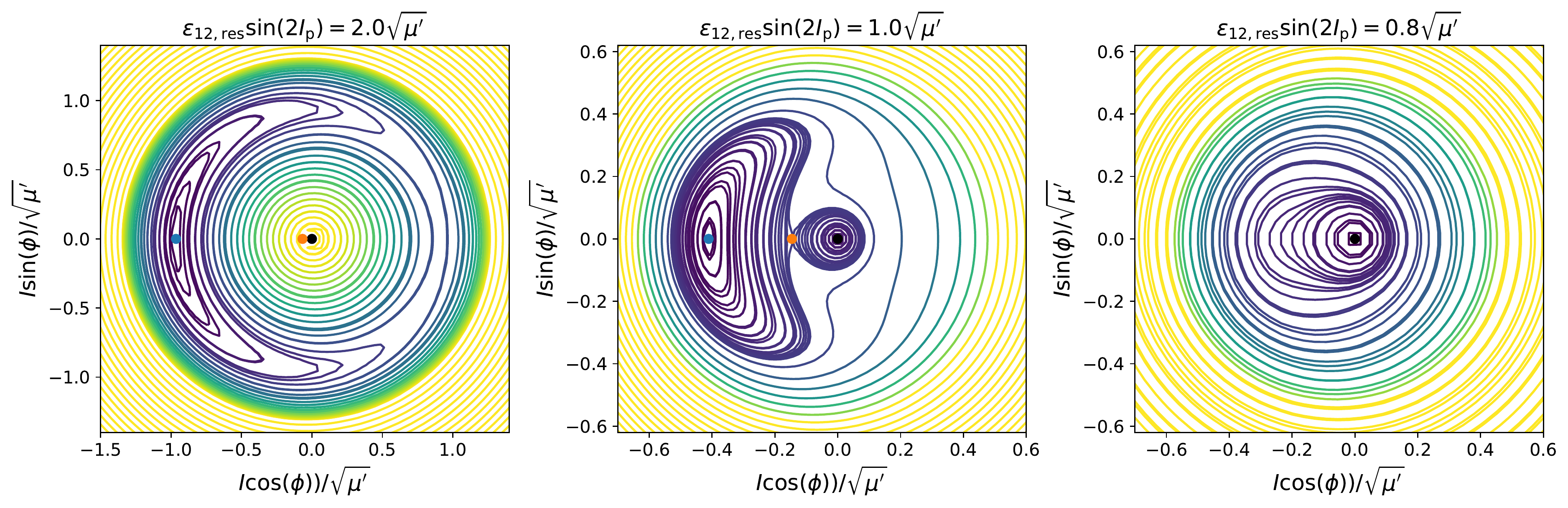}
	\caption{Hamiltonian map in the polar plane of $(I/\sqrt{\mu'}, \phi)$ for three different configurations around the stability limit of the $1:3$ MMR (Eq.~\ref{eq:stabilityapprox}). We fixed $\Omega = 0$ and $\alpha_0 = a'(1+jI_\mathrm{forced}^2/2)/a$ (Eq.~\ref{eq:alpha0}), where $a$ is the equilibrium semi-major axis corresponding to Eq.~\ref{eq:aeq}. The blue dot indicates an equilibrium, and the orange dot a saddle point. The equilibrium and saddle points merge for $\varepsilon_{12,\mathrm{res}} \sin(2 I_{\rm p}) < \sqrt{\mu'}$, swallowing the libration zone. In that case, the system is not formally in an MMR configuration, and the relevant coupling parameter is $\varepsilon_{12}$ instead of $\varepsilon_{12,\mathrm{res}}$.} \label{fig:librationdiagram}
\end{figure*}

Now let us consider the validity of the ``decoupling'' assumption, that the inner planets remain in resonance throughout the precession cycles. From Eq.~\eqref{eq:dadt}, it is clear that the only possible equilibrium of the resonance is for $\sin\phi = 0$, similar to the non-perturbed MMR. In the latter as in our case, the equilibrium at $\phi = 0$ is unstable, which is revealed by a stability analysis around that point. Let us do a similar analysis for the $\phi = \upi$ case, by studying the motion close to the precession solution, which we denote by $I_\mathrm{prec}$ and $\Omega_\mathrm{prec}$. Their expressions are given by Eq.~\eqref{eq:secular} and satisfy Eqs.~\eqref{eq:didtsecular}--\eqref{eq:dOdtsecular}. Moreover, we suppose that the longitude of the node circulates with a precession frequency roughly equal to $\omega_\mathrm{prec}$. We slightly perturb the cycle as follows:
\begin{align}
a &= a'\left(\frac{j-2}{j}- \frac{2\omega_{\mathrm{res}}}{jn'}\right)^{-\frac{2}{3}} + \delta a, \label{eq:aeq}\\
I &= I_\mathrm{prec} + \delta I,\\
\phi &= \upi + \delta\phi,\\
\Omega &= \Omega_\mathrm{prec} + \delta\Omega.\label{eq:Oeq}
\end{align}
After some algebra, the differential equations Eqs.~\eqref{eq:dadt}--\eqref{eq:dOdt} reduce to:
\begin{align}
\frac{1}{\omega_{\mathrm{res}}}\dv[2]{\delta\phi}{t} =& \left(-\frac{3}{2}n'j(j-2)I^2_\mathrm{prec}+\frac{\Delta\omega_{\rm p}\sin(2I_{\rm p})}{I_\mathrm{prec}}\cos\Omega_\mathrm{prec}\right) \delta\phi\nonumber\\
& - \frac{\omega_{\mathrm{prec}}}{\omega_{\mathrm{res}}}\frac{ \Delta\omega_{\rm p}\sin(2I_{\rm p})}{I_\mathrm{prec}}\sin\Omega_\mathrm{prec}, \label{eq:d2thetadt2} 
\end{align}
where we have neglected the dependence of the frequencies on $a$. The stability is guaranteed only if the factor in front of $\delta\phi$ is negative, and if the perturbation term proportional to $\sin\Omega$ is negligible. This gives the condition:
\begin{align}
\frac{\omega_{\mathrm{prec}}}{\omega_{\mathrm{res}}}\sin(2I_{\rm p})\frac{\Delta\omega_{\rm p} }{n'} < \frac{3}{2}j(j-2) I_\mathrm{prec}^3 \quad\text{(stability condition)}\label{eq:stability}.
\end{align}
In Figure \ref{fig:example_unstable} we show an example where Eq.~\eqref{eq:stability} is not fulfilled and the resonant angle is not librating. Thus, the secular solution Eq.~(\ref{eq:secular}) is no longer valid. The resonant angle circulates faster than the precession, so that it can be averaged in Eqs.~\eqref{eq:dadt}--\eqref{eq:dOdt}, and we get back to the non-resonant problem. The inclination angle and longitude of the node then approximately follow the pure secular behaviour described in \cite{2017AJ....153...42L}.

If the condition (\ref{eq:stability}) is not satisfied, then the stability is not guaranteed, but can nevertheless occur, depending on the entanglement between $\phi$ and $\Omega$. We will examine more thoroughly this possibility in the next subsection, in the particular case where the problem is close to its fixed point.

\subsection{Equilibrium around the fixed point}

In the previous subsection, we have examined the case where we could effectively decouple the precession from the libration of the resonant angle ($\dot\Omega \approx \omega_\mathrm{prec}$). We found that this assumption is valid for inclinations large enough to resist the perturbation from the outer perturber (Eq.~\ref{eq:stability}). Here, we examine the full problem without such assumptions, but close to the only stable fixed point. It is given by:
\begin{align}
a_{\mathrm{eq}} &= a'\left(\frac{j-2}{j} + \frac{2}{jn'}\left(\omega_{\mathrm{prec}}-\omega_{\mathrm{res}}\right)\right)^{-\frac{2}{3}},\label{eq:aeq_outer}\\
I_{\mathrm{eq}} &= I_\mathrm{forced},\\
\phi_{\mathrm{eq}} &= \upi,\\
\Omega_{\mathrm{eq}} &= 0 \label{eq:Oeq_outer}.
\end{align}
A stability analysis can be performed rigorously by deriving the eigenvalues of the $4\times4$ matrix corresponding to the evolution equations linearized near the fixed point (see Appendix \ref{sec:stability}). It reveals that the onset of instability occurs when
\begin{equation}
	\varepsilon_{12,\mathrm{res}}\sin(2I_{\rm p}) \lesssim \sqrt{\mu'}. \label{eq:stabilityapprox}
\end{equation}
This stability condition is roughly equivalent to Eq.~\eqref{eq:stability} in the strong coupling regime, where $I_\mathrm{prec}$ is set to $I_\mathrm{forced}$ and $\omega_\mathrm{prec} \sim \omega_\mathrm{res} \sim \mu' n'$. Fig.~\ref{fig:librationdiagram} shows the parameter space $(\phi, I)$, describing the MMR at fixed $\Omega$ around the stability threshold.

The stability analysis also reveals that the stability can be restored for even smaller $\varepsilon_{12,\mathrm{res}}\sin(2I_{\rm p})$. In that case, the precession frequency can be comparable to the resonant frequency, but the strong coupling between $\phi$ and $\Omega$ enables the libration to survive. An example of such evolution is shown in Fig.~\ref{fig:example_stabilityisland}. Nevertheless, this island of stability below the stability threshold Eq.~\eqref{eq:stabilityapprox} exists only close to the forced inclination, and is thus easily destroyed.

\subsection{Results for mutual inclination}

\begin{figure*}
	\centering
	\includegraphics[width=0.45\linewidth]{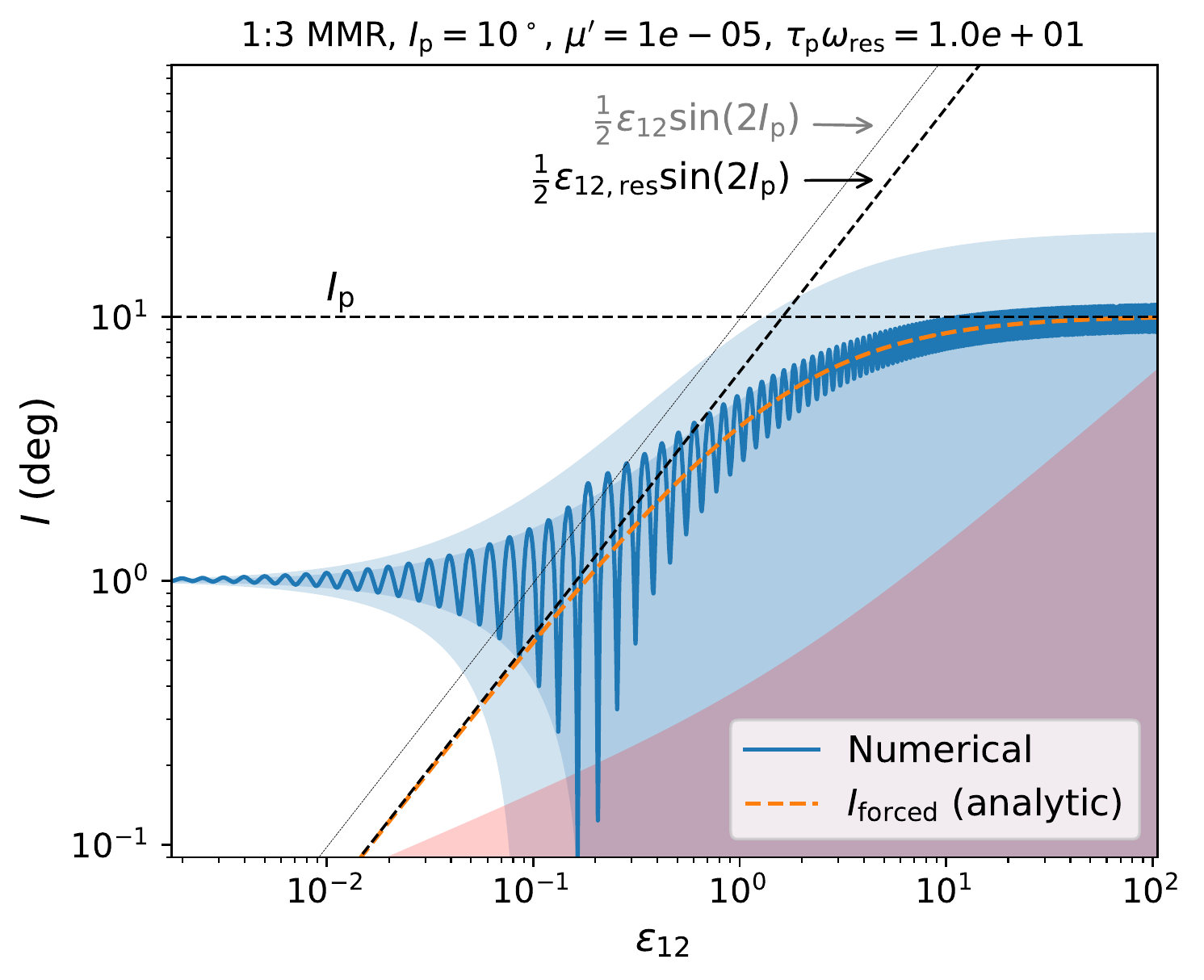}
	\includegraphics[width=0.45\linewidth]{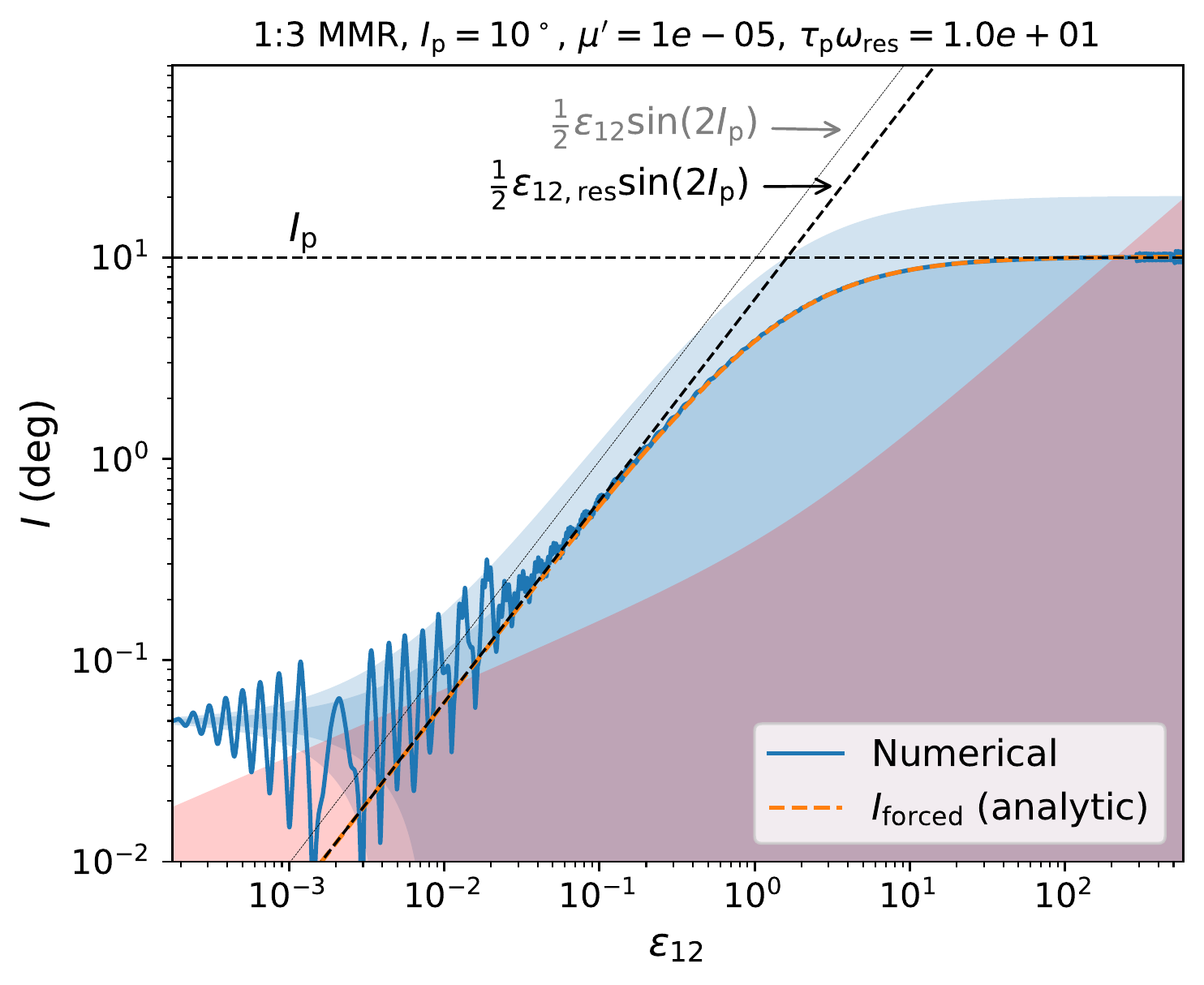}
	\caption{Inclination as a function of $\varepsilon_{12}$ with $I_0 \equiv I(t=0) = 1^\circ$ (left) or $0.05^\circ$ (right). In this calculation, the value of $\varepsilon_{12}$ increases exponentially with a characteristic time $\tau_{\rm p} \omega_{\mathrm{res}} = 10$. The orange dashed line represents the forced inclination $I_\mathrm{forced}$ (Eq.~\ref{eq:iforced}), the dark dashed line represents the forced inclination limit, and the light gray line the forced inclination limit without resonance. The blue shade represents the envelope if the system is in resonance (the dark blue corresponds to $|I-I_0| <I_\mathrm{forced}$, and the light blue corresponds to $|I-I_0| < 2I_\mathrm{forced}$). The red zone corresponds to the theoretical instability of the resonance (Eq.~\ref{eq:stability}). In the left plot, $I_0$ is high enough for the stability condition to always be fulfilled, which is not the case for the right plot. The corresponding temporal evolution of the latter is depicted in Fig.~\ref{fig:iforcedevolution}.} \label{fig:iforcedadiab}
\end{figure*}

We now compute the expected mutual inclination for the inner planets as a function of the coupling parameter $\varepsilon_{12}$. This can be done by doing numerical calculations starting from appropriate initial conditions (see below) at different $\varepsilon_{12}$ (Fig.~\ref{fig:iforced}), or by varying adiabatically $\varepsilon_{12}$ (Fig.~\ref{fig:iforcedadiab}). We integrate Eqs.~\eqref{eq:dadt}--\eqref{eq:dOdt} with the \textsc{odeint} function of the \textsc{python} module \textsc{SciPy}.

For the calculations depicted in Fig.~\ref{fig:iforced}, the integrations start at the theoretical equilibrium given by Eqs.~\eqref{eq:aeq_outer}--\eqref{eq:Oeq_outer}, and are then carried on for several precession periods. We estimate the forced inclination by averaging the mutual inclination over the entire integration time. We confirm the theoretical relation (see Eq.~\ref{eq:iforced}) between the forced inclination and the parameter $\varepsilon_{12,\mathrm{res}}$. In the strong coupling regime, the inclination is proportional to $\varepsilon_{12,\mathrm{res}}$, whereas in the weak coupling regime, the inclination reaches $I_{\rm p}$, the perturber's misalignment. The strong coupling regime of the purely secular case is also represented in Fig.~\ref{fig:iforced}, where the inclination is proportional to $\varepsilon_{12}$. The difference between the resonant and non-resonant cases is
\begin{align}
	\frac{\varepsilon_\mathrm{12,res}}{\varepsilon_{12}} = \frac{\omega_{21}}{\omega_{21,\mathrm{res}} + \omega_{21}}.\label{eq:eps12ratio}
\end{align}
For the $1:3$ MMR, this factor is approximately equal to $1/2$. Thus, the coupling between the inner planets in the resonant case is about a factor of $2$ larger than in the purely secular case: all other things being equal, it is harder to increase the mutual inclination. Finally, we retrieve the stability condition around the fixed point (Eq.~\ref{eq:instability}): most of the time, the stability of the resonance is ensured, even for $I \approx I_{\rm p}$ and large $\varepsilon_{12,\mathrm{res}}$.

\begin{figure}
	\centering
	\includegraphics[width=\linewidth]{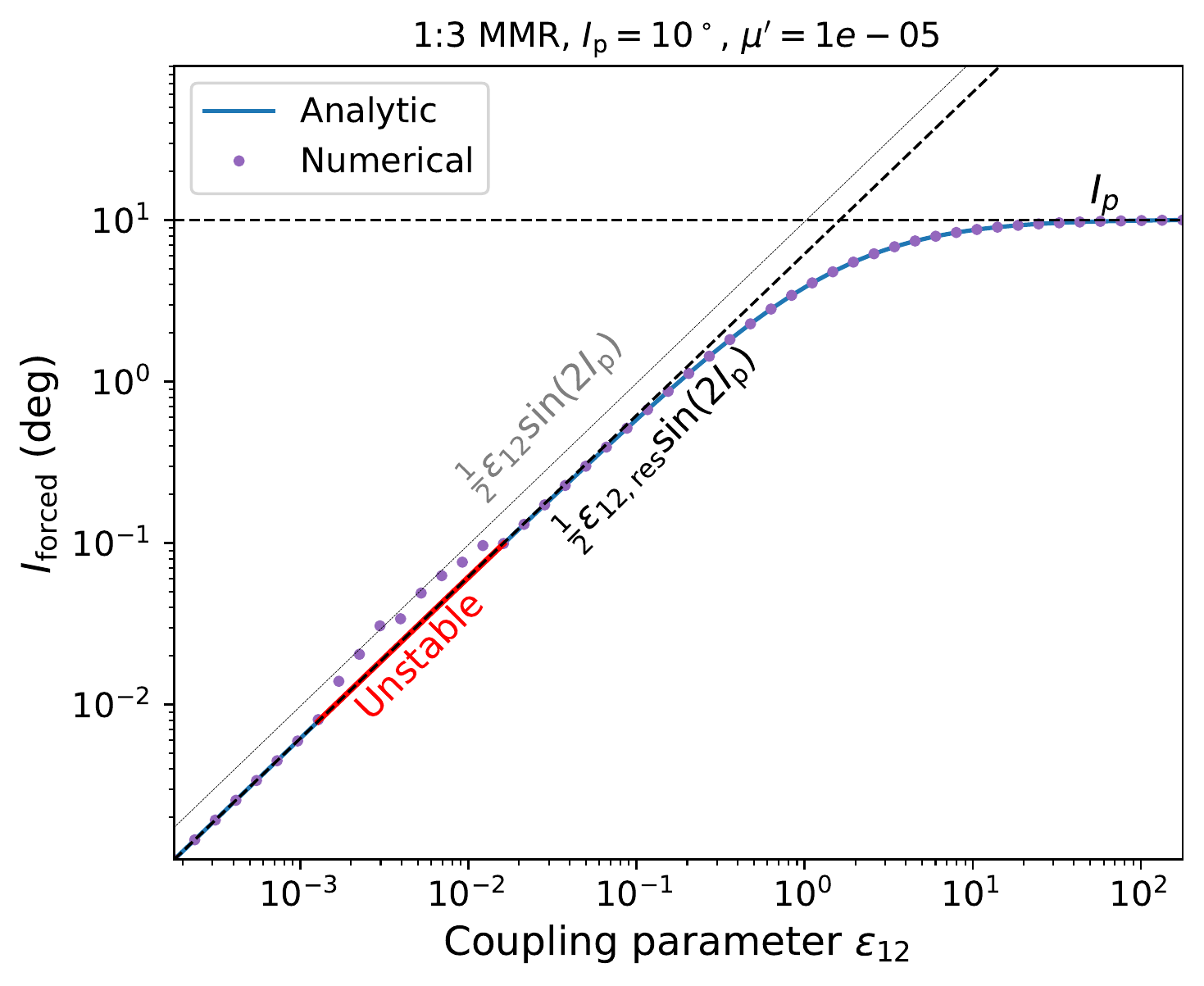}
	\caption{Forced mutual inclination of two inner planets in $1:3$ MMR as a function of $\varepsilon_{12}$, in the case where $m_2$ is a test mass. The external perturber is inclined by $I_{\rm p} = 10^\circ$. Each purple point represents the average inclination of an integration with initial conditions close to the theoretical equilibrium (Eqs.~\ref{eq:aeq_outer}--\ref{eq:Oeq_outer}), while the blue line represents the theoretical forced inclination (Eq.~\ref{eq:iforced}). The red line corresponds to the theoretical instability derived in Appendix A (Eq.~\ref{eq:instability}) and described in Sec.~3.3. The dark dashed line represents the forced inclination limit in the resonant case (Eq.~\ref{eq:iforced}), and the light gray line represents the forced inclination limit in the pure secular case \citep{2017AJ....153...42L}.} \label{fig:iforced}
\end{figure}

\begin{figure}
	\centering
	\includegraphics[width=\linewidth]{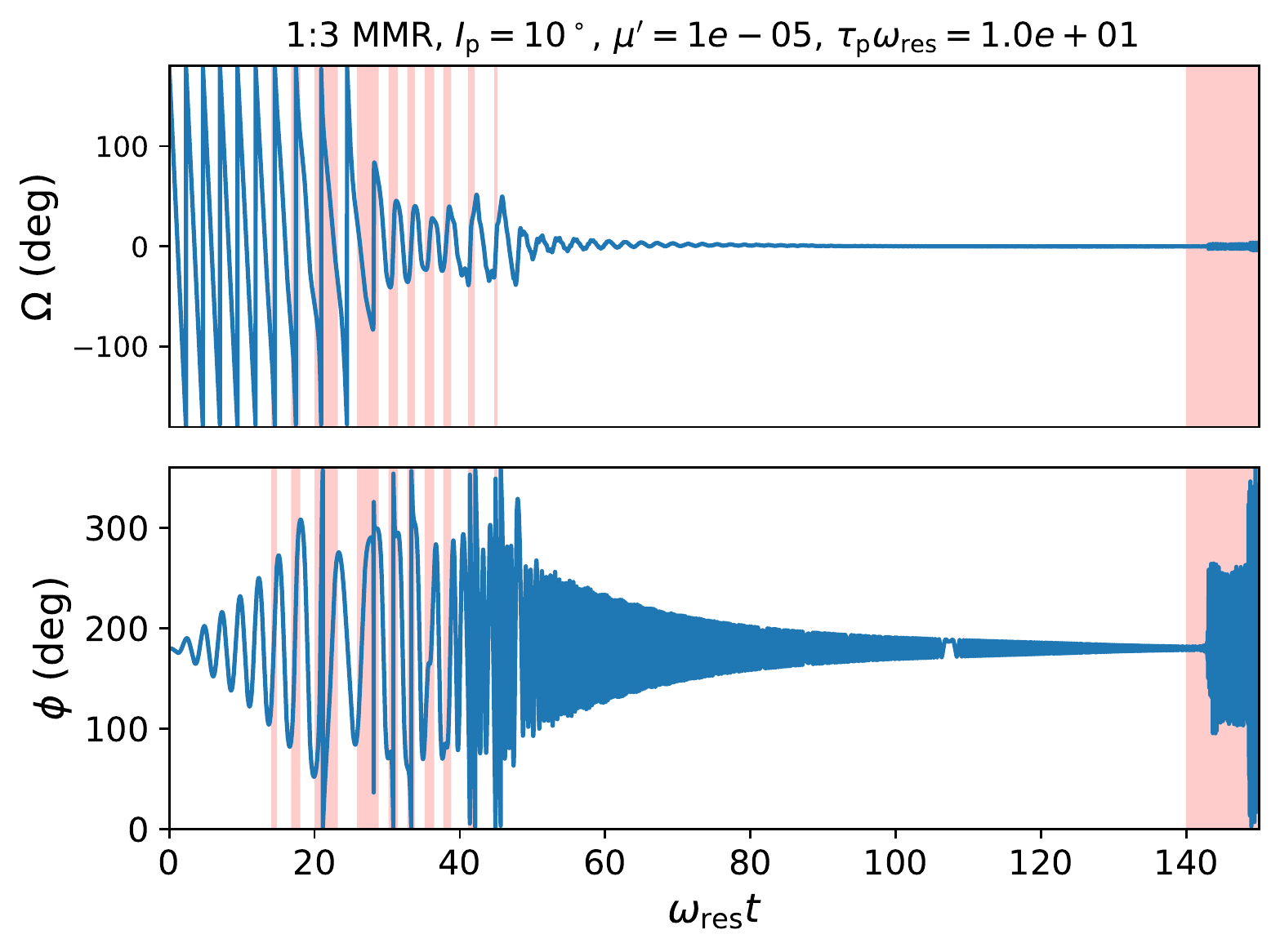}
	\caption{Time evolution of the longitude of the node $\Omega$ and the resonant angle $\phi$, corresponding to the right-hand panel of Fig.~\ref{fig:iforcedadiab}, where $\varepsilon_{12}$ increases exponentially. In the beginning, the inner planet pair is in resonance (the perturber has a negligible impact). Then, the perturber has enough influence to break the resonance (red zones, Eq.~\ref{eq:stability} is not satisfied). When the inclination gets forced to higher values as $\varepsilon_{12}$ increases (Eq.~\ref{eq:iforced}), the stability is restored. Finally, the inclination stabilizes around $I_{\rm p}$ while the perturber keeps growing. Again, the stability condition  Eq.~\eqref{eq:stability} does not hold and the resonant angle circulates.} \label{fig:iforcedevolution}
\end{figure}

In Fig.~\ref{fig:iforcedadiab}, the system is set initially at equilibrium (Eqs.~\ref{eq:aeq}--\ref{eq:Oeq}) and the outer perturber has a negligible effect ($\varepsilon_{12} < 10^{-3}$). We suppose that the inner planets have a primordial misalignment $I_0 \equiv I(t=0)$ given by their formation and early evolution history. The strength of the perturber is then gradually increased, on a timescale much larger than the precession period. This is equivalent to an outer perturber migrating inwards. The system is not responding significantly to the perturber's influence until the initial and the forced inclinations have the same magnitude: the longitude of the node then begins to librate around $0^\circ$ and the inclination remains close to the forced inclination, thus increasing with $\varepsilon_{12,\mathrm{res}}$. The resonance might hold during the entire process (i.e. the resonant angle keeps librating), depending on whether the condition for stability (\ref{eq:stability}) is fulfilled when $I \approx I_\mathrm{forced}(\varepsilon_{12,\mathrm{res}})$. If it is not (see the right-hand panel of Fig.~\ref{fig:iforcedadiab}), then the resonance will temporarily break and the forced inclination will correspond to the non-resonant/purely secular case studied in \cite{2017AJ....153...42L}. The corresponding evolution for the resonant angle and the longitude of the node is represented in Fig.~\ref{fig:iforcedevolution}.

\section{Test mass-planet resonance perturbed by an external body: inner test mass}
\label{sec:testmass-planet_inner}

Here we present the case where planet 1 (the innermost planet) is a test mass ($m_1 \ll m_2, m_{\rm p}$). As in Section \ref{sec:testmass-planet}, we adopt the notations where primed quantities refer to the finite-mass planet and unprimed quantities to the test mass: $(a_1=a) < (a_2=a') < a_{\rm p}$. We consider the problem in the orbital plane of the planet, so that Fig.~\ref{fig:notationsvectors} still applies, except $\vec{\hat{z}}= \vec{\hat l_2 }$ and $\vec{\hat{l}} = \vec{\hat l_1}$.

\subsection{Hamiltonian and evolution equations}

As in Section \ref{sec:testmass-planet}, the Hamiltonian per unit mass is a sum of several terms. To the order 2 in $I$, the interaction with the other planet and the central star is
\begin{equation}
H_{12} = -\frac{GM}{2a} - \frac{1}{2}\omega_\mathrm{res} \Lambda I^2\cos\phi + \frac{1}{2}  \omega_{12} \Lambda I^2.
\end{equation}
where $\omega_{\mathrm{res}}$ now denotes $\omega_{12, \mathrm{res}}$.The perturbation from the outer body writes
\begin{equation}
H_{1p} = \frac{1}{2}\omega_\mathrm{1{\rm p}} \Lambda \sin^2 \theta_{\rm p},
\end{equation}
where $\theta_{\rm p}$ is the relative angle between the orbital planes of $m$ and $m_{\rm p}$ (Eq.~\ref{eq:thetap}). The term associated with the rotating frame (the orbital plane of planet 2) is:
\begin{equation}
H_\mathrm{rot} = (\omega_{2{\rm p}}\cos I_{\rm p}) \vec{\hat l_{\rm p}}.\frac{\vec{L_1}}{m} =  \omega_{2{\rm p}}\Lambda\cos I_{\rm p} \cos\theta_{\rm p}.
\end{equation}
Finally, when making a canonical transformation from $\lambda$, $-\Omega$ to $\phi$, $\Omega$, an additional term must be added (see Section \ref{sec:testmass-planet}):
\begin{equation}
H_S = \frac{j}{j-2}n'\Lambda.
\end{equation}

Rewriting the Hamilton-Jacobi equations to the first orders in $I$, $\mu'$ and $\mu_{\rm p}$ (see Section \ref{sec:testmass-planet}), we obtain:
\begin{align}
\dv{a}{t} &= (j-2) \omega_{\mathrm{res}} a I^2 \sin\phi\label{eq:dadt_inner},\\
\dv{I}{t} &=  -\omega_{\mathrm{res}} I \sin\phi - \frac{\Delta\omega_{\rm p}}{2} \sin(2I_{\rm p}) \sin\Omega \label{eq:didt_inner},\\
\dv{\phi}{t} &= -\Delta n - 2\omega_{\mathrm{res}} \cos\phi+ \frac{\Delta\omega_{\rm p}}{I}\sin(2I_{\rm p})\cos\Omega\label{eq:dphidt_inner}, \\
\dv{\Omega}{t} &= \omega_{\mathrm{res}}\cos\phi -\omega_{12} + \Delta\omega_{\rm p}\cos ^2 I_{\rm p} - \frac{\Delta\omega_{\rm p}}{2I}\sin(2I_{\rm p})\cos\Omega\label{eq:dOdt_inner},
\end{align}

\noindent where $\Delta n$ characterizes the distance from the nominal location of the MMR (Eq.~\ref{eq:eta}). This system of differential equations is similar to the outer test mass case, only the coefficient in front of the semi-major axis and the signs in front of the $\Delta\omega_{\rm p}$ terms are different.

\subsection{Forced inclination and MMR stability}

From Equations \eqref{eq:dadt_inner}--\eqref{eq:dOdt_inner}, we define the ``total'' precession frequency $\omega_{\mathrm{prec}}$ of the test mass in the rotating frame as:
\begin{equation}
\omega_{\mathrm{prec}} \equiv  |\omega_{\mathrm{res}} +\omega_{12} -\left(\omega_{2{\rm p}}-\omega_{1{\rm p}}\right)\cos ^2 I_{\rm p}|.\label{eq:wprec_inner}
\end{equation}

Following the approach of Section \ref{sec:testmass-planet}.2, we can solve Eqs.~\eqref{eq:dadt_inner}--\eqref{eq:dOdt_inner} exactly by assuming that the resonant angle $\phi$ stays close to $\upi$. The forced inclination retains the same expression as Eq.~\ref{eq:iforced}.

\begin{figure}
	\centering
	\includegraphics[width=\linewidth]{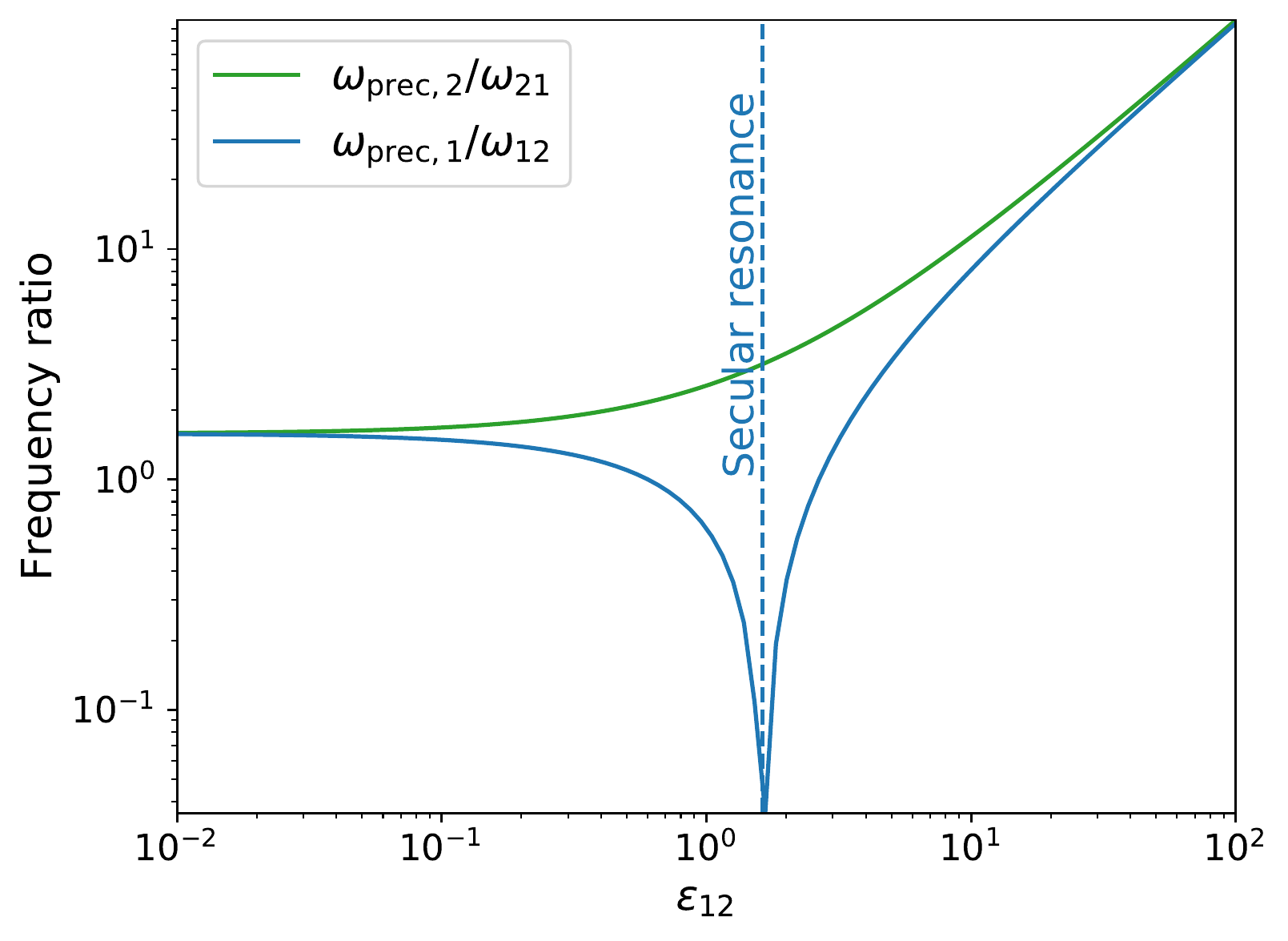}
	\caption{True precession frequency of the test mass as a function of the precession frequency induced by the fellow planet, for a perturber misalignment $I_{\rm p} = 10^\circ$ and a pair of inner planets in 1:3 MMR ($\omega_{\mathrm{prec},2}$ corresponds to the outer test mass case, $\omega_{\mathrm{prec},1}$ to the inner test mass case). For the inner test mass case (blue), the true precession frequency cancels as $\omega_{12}+\omega_{\mathrm{res}} = \Delta\omega_{\rm p} \cos^2 I_{\rm p}$, leading to a significant increase in the relative inclination $I$.} \label{fig:frequencies}
\end{figure}

A comparison between the precession frequencies for the outer test mass case (Eq.~\ref{eq:wprec}, $\omega_\mathrm{prec,2}$) and inner test mass case (Eq.~\ref{eq:wprec_inner}, $\omega_\mathrm{prec,1}$) is represented in Figure \ref{fig:frequencies}. As it appears clearly in both the plot and Eq.~\eqref{eq:wprec_inner}, $\omega_{\mathrm{prec}}$ can be null in the inner test mass case. This is an indication of a secular inclination resonance. In this case, the forced inclination formally diverges (see Eq.~\ref{eq:iforced}). Since our analytic theory breaks down  for large inclination, we will not dwell on the study of the cases close to $\omega_{\mathrm{prec}} = 0$.

The stability of the MMR, and thus the validity of the assumption $\phi \approx \upi$, leads to the same constraints in this case as in the outer test mass case. By assuming $\dot\Omega \approx \omega_\mathrm{prec}$ and performing a stability analysis around the precessing solutions, we recover the stability condition given by Eq.~\eqref{eq:stability}. On the other hand, a stability analysis can be performed as in Section \ref{sec:testmass-planet}.4 around the fixed point, without any assumptions on $\dot\Omega$. In the inner test mass case, the fixed point is
\begin{align}
a_{\mathrm{eq}} &= a'\left(\frac{j}{j-2} + \frac{2}{jn'}\left(\omega_\mathrm{res}+\omega_{\mathrm{prec}}\cos\Omega_{\mathrm{eq}}\right)\right)^{-\frac{2}{3}},\label{eq:aeq_inner}\\
I_{\mathrm{eq}} &= I_\mathrm{forced},\\
\phi_{\mathrm{eq}} &= \upi,\\
\Omega_{\mathrm{eq}} &= \upi \text{ if }\omega_{12}+\omega_{\mathrm{res}} > \Delta\omega_{\rm p} \cos^2 I_{\rm p}\text{, 0 otherwise}\label{eq:Oeq_inner}.
\end{align}
The condition for stability Eq.~\eqref{eq:instability} and its approximation Eq.~\eqref{eq:stabilityapprox} remain valid.

\subsection{Results for mutual inclination}

\begin{figure}
	\centering
	\includegraphics[width=\linewidth]{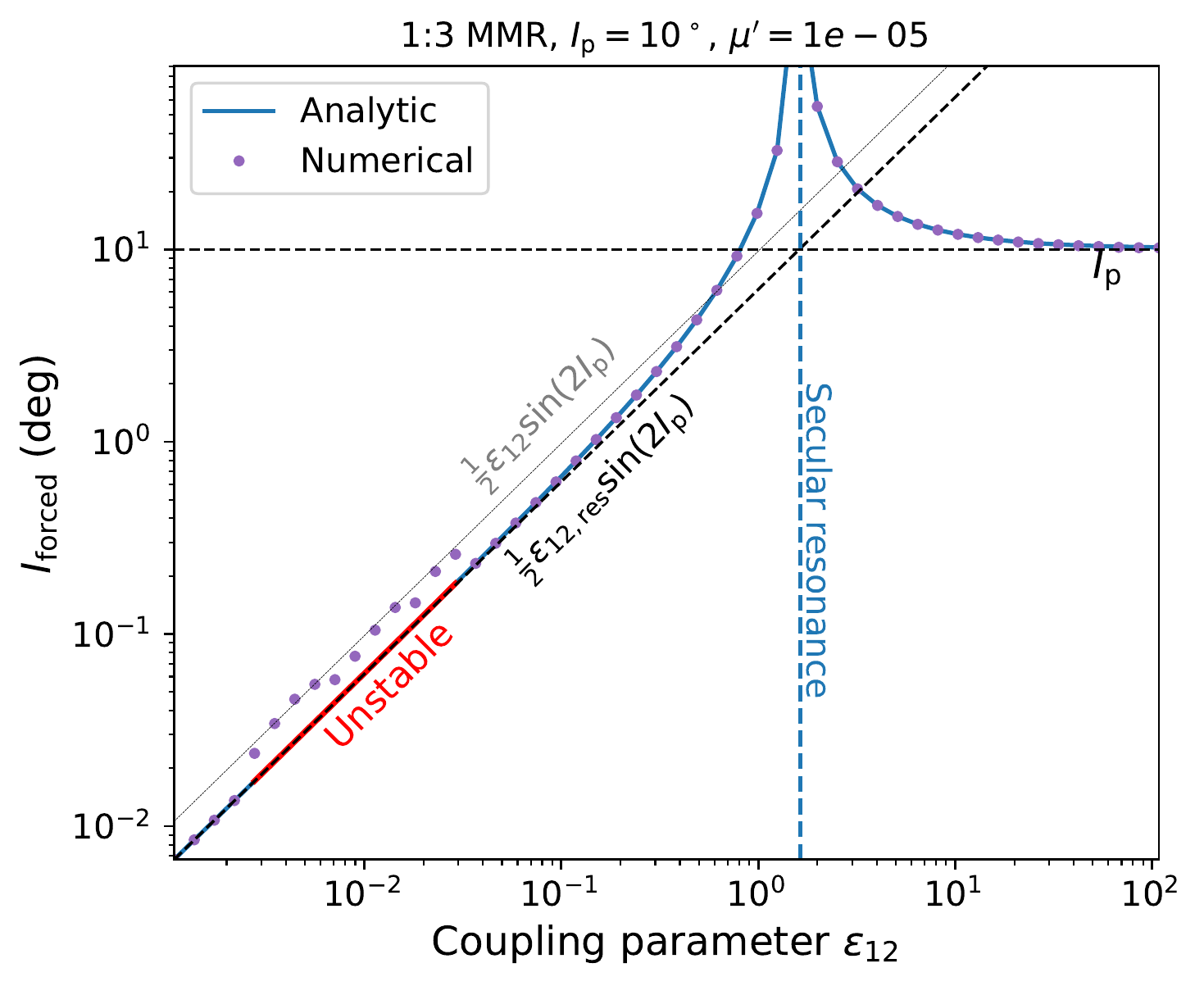}
	\caption{Forced mutual inclination as a function of $\varepsilon_{12}$, for configurations identical to Fig.~\ref{fig:iforced} except that $m_1$ is now the test mass. Each purple point represents the average inclination for an integration with initial conditions close to equilibrium (Eqs.~\ref{eq:aeq_inner}--\ref{eq:Oeq_inner}), and the blue line represents the theoretical forced inclination (Eq.~\ref{eq:iforced}). The red line corresponds to the theoretical instability derived in the Appendix (Eq.~\ref{eq:instability}) and described in Sec.~3.3 and 4.2. The dark dashed line represents the forced inclination limit assuming resonance ($\phi \approx \upi$, Eq.~\ref{eq:iforced}), and the light gray one represents the forced inclination limit in the purely secular case. The equations are not valid for large inclinations close to the secular resonance ($\varepsilon_{12} \sim 1$).} \label{fig:ieq}
\end{figure}

Again, there are different ways to compute the expected mutual inclination for this problem as a function of the coupling parameter $\varepsilon_{12}$. Varying adiabatically $\varepsilon_{12}$ is compromised here because of the divergence at $\omega_{\mathrm{prec}} = 0$, corresponding to $\varepsilon_{12} = (1+\omega_{\mathrm{res}}/\omega_{12})/\cos^2(I_{\rm p})$ $\approx$ 1. We instead integrate the system of equations Eqs.~\eqref{eq:dadt_inner}--\eqref{eq:dOdt_inner} for different $\varepsilon_{12}$, close to the theoretical equilibrium Eqs.~\eqref{eq:aeq_inner}--\eqref{eq:Oeq_inner}, for several precession periods, and average the inclination to derive the forced inclination. The results are shown in Fig.~\ref{fig:ieq}. We confirm that the forced inclination is proportional to $\varepsilon_{12,\mathrm{res}}$ for strong couplings as given in Eq.~\eqref{eq:iforced}, when the stability criterion Eq.~\eqref{eq:instability} is fulfilled.

\section{Planet-planet resonance perturbed by an external body} \label{sec:planet-planet}

When both inner planets have finite masses, the inclination dynamics become more complex. In the plane of reference (to be specified below), let $I_1$ and $I_2$ be the inclinations of the two planets, and $\Omega_1$ and $\Omega_2$ the longitudes of the node. We define the orientation of the reference plane so that the external perturber has $\Omega_{\rm p}=0$. The mutual inclination between the two inner planets is given by:
\begin{equation}
\cos I = \cos I_1 \cos I_2 + \cos(\Omega_2-\Omega_1) \sin I_1 \sin I_2.
\end{equation}

\subsection{Strong coupling limit}

In the strong coupling limit ($\varepsilon_{12}\ll 1$), the inner planets remain nearly coplanar with small mutual inclination. Let ${\vec L}={\vec L}_1+ {\bf L}_2\equiv L {\vec {\hat l}}$ be the total angular momentum of the two inner planets, with $L\simeq L_1+L_2$. The total torque on the inner planets from the external perturber is
\begin{align}
\omega_{1{\rm p}}\cos\theta_{1{\rm p}}\vec{L_1}\times \vec{\hat l_{\rm p}} + \omega_{2{\rm p}}\cos\theta_{2{\rm p}}\vec{L_2}\times \vec{\hat l_{\rm p}}.
\end{align}
With $\vec {\hat l_1}$ nearly parallel to $\vec {{\hat l}_2}$, we find
\begin{align}
\dv{\vec{L}}{t} \simeq \omega_{\rm p} (\vec {\hat l} \cdot \vec {\hat l_{\rm p}}) {\bf L}\times {\vec {\hat l}}_{\rm p},
\end{align}
where
\begin{equation}
\omega_{\rm p} = \frac{\omega_{1{\rm p}}L_1+\omega_{2{\rm p}}L_2}{L}\label{eq:wp}.
\end{equation}
Thus, $\vec {\hat l}$ rotates around $\vec{ \hat l}_{\rm p}$ at the rate $-\omega_{\rm p} \vec{\hat l}\cdot \vec{\hat l_{\rm p}}$. This rate is approximately constant for $\varepsilon_{12} \ll 1$, or $\mu_1$ or $\mu_2 \ll 1$ (Appendix \ref{sec:rotatingframe}). To examine the mutual inclination between the two inner planets, it is convenient to work in this rotating frame in the strong coupling regime, where we have the following relations:
\begin{align}
&I = I_1+I_2,\\
&L_1\sin I_1 = L_2\sin I_2,\\
&\Omega_1 = \Omega_2 + \upi.
\end{align}
Thus, a small mutual inclination $I$ implies that both $I_1$ and $I_2$ are small. 

\begin{figure*}
	\centering
	\includegraphics[width=0.8\linewidth]{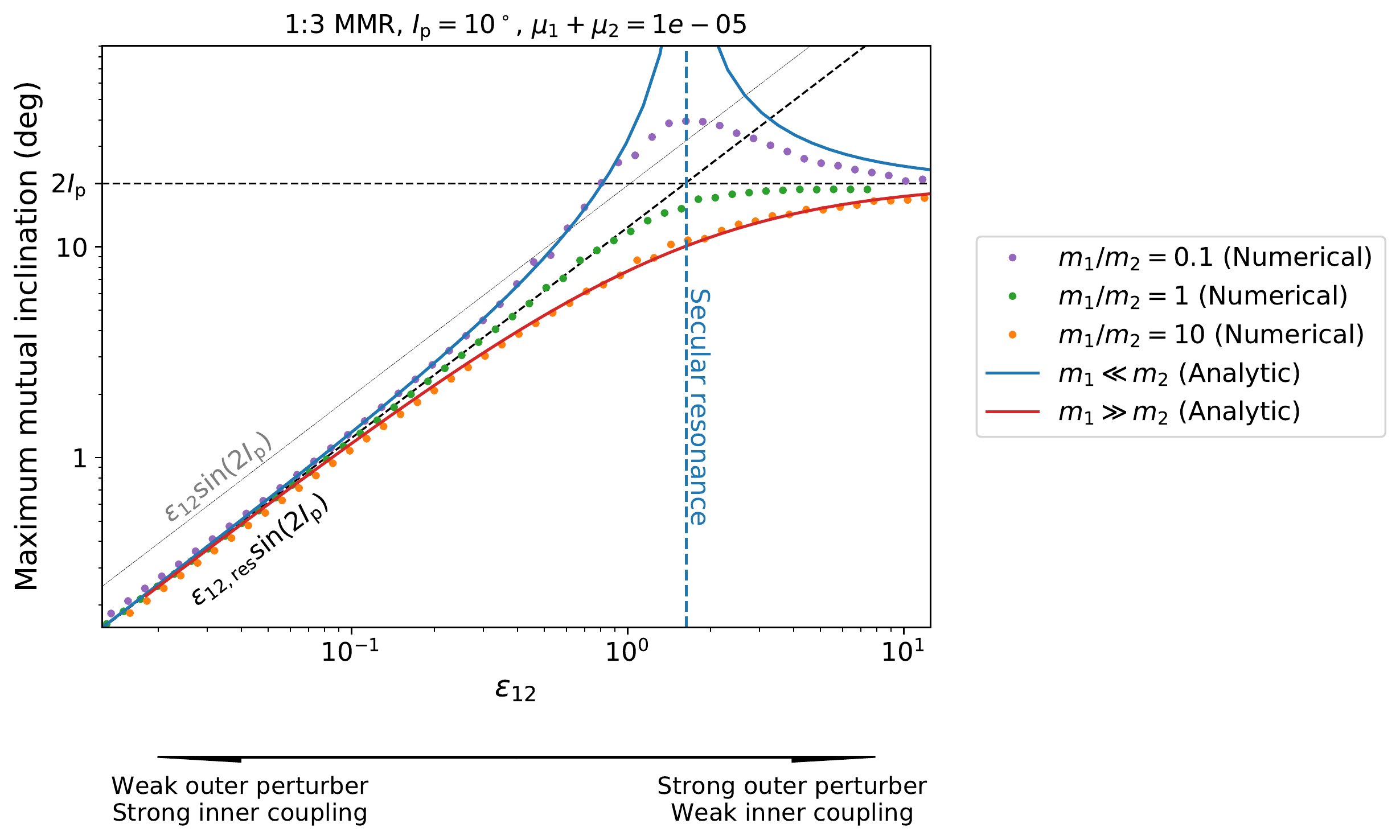}
	\caption{Maximum mutual inclination amplitude of the inner planets as a function of $\varepsilon_{12}$, for different mass ratios $m_1/m_2$. The pair of inner planets is in the $1:3$ MMR, and the external perturber is inclined by $I_p=10^\circ$. Each point represents an integration, and the blue and red lines represent the theoretical maximum inclination $2I_\mathrm{forced}$ (Eq.~\ref{eq:imax}) in the test mass cases. The dark dashed line represents the forced inclination limit in the strong coupling regime (Eq.~\ref{eq:imax_comparablemass}), and the light gray line represents the same limit in the purely secular problem. The equations are not valid for large inclinations close to the secular resonance.} \label{fig:imax}
\end{figure*}

We now write down the Hamiltonian for the two inner planets. 
First, to the order 2 in $I$, the secular interactions between the inner planets and with the central star give \citep{2000ssd..book.....M}:
\begin{align}
H_{12/21} &= -\frac{GMm_1}{2a_1} -\frac{GMm_2}{2a_2} + \frac{1}{2}  \omega_{12} L_1 I^2,\label{eq:H1221}
\end{align}
(note that $\omega_{12}L_1 = \omega_{21}L_2$). Second, the resonant interaction is a combination of three terms, which correspond to two resonant angles as well as a combination of the two:
\begin{align}
\phi_1 &= j\lambda_2 - (j-2)\lambda_1 - 2 \Omega_1,\\
\phi_2 &= j\lambda_2 - (j-2)\lambda_1 - 2 \Omega_2,\\
H_{\mathrm{res}} &= - \frac{1}{2}\omega_\mathrm{12,res} L_1 I_1^2\cos\phi_1- \frac{1}{2}\omega_\mathrm{21,res} L_2 I_2^2\cos\phi_2\nonumber\\
& + \omega_\mathrm{12,res} L_1 I_1 I_2\cos(\frac{\phi_1+\phi_2}{2}).\label{eq:Hres}
\end{align}
The secular interaction on $m_1$ and $m_2$ from $m_p$ gives:
\begin{align}
H_{1{\rm p}} &= \frac{1}{2} \omega_{1{\rm p}} L_1 \sin^2 \theta_{1{\rm p}},\\
H_{2{\rm p}} &= \frac{1}{2} \omega_{2{\rm p}} L_2 \sin^2 \theta_{2{\rm p}},\label{eq:H2p}
\end{align}
with
\begin{align}
\cos\theta_{1{\rm p}} &= \cos I_1 \cos I_{\rm p}  + \cos\Omega_1\sin I_1 \sin I_{\rm p}, \\
\cos\theta_{2{\rm p}} &= \cos I_2 \cos I_{\rm p} + \cos\Omega_2\sin I_2\sin I_{\rm p}.
\end{align} 
Finally, due to the rotation of our reference plane, we add a ``rotational'' term to the Hamiltonian \citep{2014AmJPh..82..769T}:
\begin{align}
H_\mathrm{rot} &= (\omega_{\rm p}\cos I_{\rm p}) \vec{\hat l_{\rm p}}.(\vec{L_1}+\vec{L_2})\nonumber\\ 
&=  \omega_{\rm p}\cos I_{\rm p} (\cos\theta_{1{\rm p}} L_1 + \cos\theta_{2{\rm p}} L_2).
\end{align}
The Hamiltonian of the two inner planets in the rotating frame is thus:
\begin{equation}
H = H_{12/21} + H_{\mathrm{res}} + H_{1{\rm p}} + H_{2{\rm p}} + H_\mathrm{rot}.
\end{equation}
It has four angle-action pairs:
\begin{align}
\lambda_1&,~ L_1 = m_1 \sqrt{GMa_1},\\
\lambda_2&,~ L_2 = m_2\sqrt{GMa_2},\\
-\Omega_1&,~ Z_1\approx L_1 I_1^2/2,\\
-\Omega_2&,~ Z_2\approx L_2 I_2^2/2.
\end{align}
We can then derive the eight Hamilton-Jacobi equations (Appendix \ref{sec:appendixcomparablemass}). We notice that the Hamiltonian only depends on $\lambda_1$ and $\lambda_2$ through $j\lambda_2-(j-2)\lambda_1$. The associated conserved quantity is
\begin{equation}
J = jL_1 + (j-2)L_2,
\end{equation}
which implies that $a_1$ and $a_2$ are not independent. We will thus replace their evolution equations by the evolution of $\alpha = a_1/a_2$. The problem is now described by six differential equations.

Finally, since in our rotating frame $I_1$ is related to $I_2$, $\Omega_1$ is related to $\Omega_2$, and $\phi_1 = \phi_2 \equiv \phi$, we have (Appendix \ref{sec:appendixcomparablemass})
\begin{align}
\dv{\alpha}{t} =& \left((j-2) \omega_{\mathrm{12,res}} + j \omega_{\mathrm{21,res}}\right) \alpha I^2 \sin\phi,\\
\dv{I}{t} =& - (\omega_{\mathrm{12,res}}+\omega_{\mathrm{21,res}}) I \sin\phi -  \frac{\Delta\omega_{\rm p}}{2}\sin(2I_{\rm p})\sin\Omega_1,\\
\dv\phi{t} =& -\Delta n - 2\left(\omega_{\mathrm{12,res}} + \omega_{\mathrm{21,res}}\right) \cos\phi\nonumber\\ & + \frac{\Delta\omega_{\rm p}}{I}\cos\Omega_1\sin(2I_{\rm p}),\\
\dv{\Omega_1}{t} =& - (\omega_{12}+\omega_{21}) +  \left(\omega_{\mathrm{12,res}} + \omega_{\mathrm{21,res}}\right) \cos\phi\nonumber\\ &- \frac{\Delta\omega_{\rm p}}{2I} \cos\Omega_1\sin (2I_{\rm p}),
\end{align}
where we still have $\Delta\omega_{\rm p} = \omega_{2{\rm p}}-\omega_{1{\rm p}}$ and $\Delta n = (j-2)n_1-jn_2$.

These equations are essentially similar to the corresponding equations in the test mass cases studied in Sections \ref{sec:testmass-planet} and \ref{sec:testmass-planet_inner} (Eqs.~\ref{eq:dadt}--\ref{eq:dOdt} and \ref{eq:dadt_inner}--\ref{eq:dOdt_inner}). The only difference is that the precession frequencies between the inner planets, $\omega_{12}$ and $\omega_{21}$, $\omega_{12, \mathrm{res}}$ and $\omega_{21,\mathrm{res}}$ are added up, strengthening their coupling. Thus, in the strong coupling limit, the forced mutual inclination and the maximum inclination amplitude are
\begin{align}
&I_\mathrm{forced} = \frac{1}{2}\varepsilon_{12, \mathrm{res}} \sin(2 I_{\rm p}), \label{eq:iforced_comparablemass}\\
&I_\mathrm{max}-I_\mathrm{min} = \varepsilon_{12, \mathrm{res}} \sin(2 I_{\rm p}),\label{eq:imax_comparablemass}
\end{align}
where $\varepsilon_{12, \mathrm{res}}$ is given by Eq.~\eqref{eq:eps12res}. Equations \eqref{eq:iforced_comparablemass}--\eqref{eq:imax_comparablemass} are valid for arbitrary $I_{\rm p}$, as long as we are in the strong coupling regime $\varepsilon_{12} \ll 1$.

\subsection{Weak coupling limit}

In the weak coupling limit, corresponding to $\varepsilon_{12} \gg 1$, the vectors $\vec{\hat l_1}$ and $\vec{\hat l_2}$ precess around $\vec{\hat l_{\rm p}}$ independently, with constant $\cos\theta_{1{\rm p}}$ and $\cos\theta_{2{\rm p}}$. We can thus expect to have
\begin{align}
I_\mathrm{max}-I_\mathrm{min} &= 2I_{\rm p}.
\end{align}
This is valid for arbitrary $I_{\rm p}$.

\subsection{General coupling, small $I_{\rm p}$}

In the general case, the total angular momentum of the inner planets $\vec L = \vec{L_1} + \vec{L_2}$ has a complex motion because its misalignment $I_{\rm p}$ with the outer perturber varies in time (see Appendix \ref{sec:rotatingframe}). In order to write down the Hamiltonian of the MMR, we have to assume that $I_1$ and $I_2$ are small. We therefore study the problem in the initial orbital plane of the inner planets, assuming they are initially coplanar or nearly coplanar---note that strict coplanarity might compromise the stability of the MMR. To ensure $I_1$, $I_2$ remain small during the evolution, we assume that the perturber inclination is not too large ($I_{\rm p} \ll 1$).

The Hamiltonian is then the same as Section \ref{sec:planet-planet}.3 (Eqs.~\ref{eq:H1221}, \ref{eq:Hres}--\ref{eq:H2p}), without the ``rotational'' term:
\begin{equation}
H = H_{12/21} + H_{\mathrm{res}} + H_{1{\rm p}} + H_{2{\rm p}}. \label{eq:hamiltonian}
\end{equation}
We derive the eight associated Hamilton-Jacobi equations in Appendix \ref{sec:appendixcomparablemass}, and the reduction to six differential equations. Those equations cannot be solved analytically, but can be integrated numerically.

Figure \ref{fig:imax} shows the maximum amplitude of the mutual inclination between the inner planets as a function of the coupling parameter $\varepsilon_{12}$, for different mass ratios $m_1/m_2$, obtained numerically. The integrations start in a nearly coplanar configuration for the pair of inner planets in the 1:3 MMR. Note that to guarantee the stability of the MMR (see Eq.~\ref{eq:stability}), the initial inclination $I_\mathrm{min}$ should not be too small; we find that $I_\mathrm{min} = 1^\circ$ ensures this for the range of coupling parameters we have tested. The integrations are then carried on for several precession periods, and the maximum mutual inclination amplitude $I_\mathrm{\max} - I_\mathrm{\min}$ is then computed. In the strong coupling regime ($\varepsilon_{12} \ll 1 $), we recover our analytical results (Eq.~\ref{eq:imax_comparablemass}) that the maximum mutual inclination is proportional to $\varepsilon_{12,\mathrm{res}}$; in the weak coupling regime, the maximum amplitude reaches $2 I_{\rm p}$. For intermediate values of $\varepsilon_{12}$, the behaviour is very similar to the test mass cases, with a smooth change between regimes for $m_2 < m_1$ and a secular resonance feature for $m_1 < m_2$, where the mutual inclination $I$ reaches very high values around $\varepsilon_{12} \sim 1$. Once again, all other things being equal, we observe that it is harder to increase the mutual inclination in a resonant configuration compared to the non-resonant case for strong coupling $\varepsilon_{12} < 1$.

\section{Summary and Discussion}
\label{sec:conclusion}

\subsection{Key results}

We have calculated the excitation of mutual inclination in an inner two-planet system due to an inclined external companion. Such a configuration resembles the observed super-Earth systems surrounded by cold Jupiters. Our results, summarized in Fig.~\ref{fig:imax}, generalize that of previous studies \citep[e.g.,][]{2017AJ....153...42L} by considering two planets in second-order mean-motion resonances (first-order MMRs do not affect mutual inclinations). We have identified a key dimensionless parameter, $\varepsilon_{12, \mathrm{res}}$, defined in Eq.~\eqref{eq:eps12res}, that characterizes the relative strength of the perturber with respect to the coupling between the two inner planets. In order of magnitude, $\varepsilon_\mathrm{12,res} \sim (m_{\rm p}/m_{1,2})(a_{1,2}/a_{\rm p})^3$, where $m_{1,2}$ and $m_{\rm p}$ are the masses of the inner planets and external perturber, and $a_{1,2}$ and $a_{\rm p}$ are their semi-major axes, respectively. In the strong coupling regime (weak perturber, small $\varepsilon_{12, \mathrm{res}}$), the induced misalignment within the inner planets is proportional to $\varepsilon_{12, \mathrm{res}}\sin(2I_{\rm p})$. In the weak coupling regime (strong perturber, large $\varepsilon_{12, \mathrm{res}}$), the maximum misalignment can reach $2 I_{\rm p}$. In the intermediate regime, when $m_1 < m_2$, a secular resonance feature might drive the mutual inclination to very high values, even for low $I_{\rm p}$. These behaviours are similar to those of pure secular systems (i.e. when the inner planets are not in MMR) studied by \cite{2017AJ....153...42L}. The difference between the resonant case and the purely-secular case lies in the coupling parameter. We have shown that the inner coupling is strengthened by the MMR, or equivalently, the outer planet's effective perturbative effect is reduced. For the $j:j-2$ second-order MMR, $\varepsilon_{12, \mathrm{res}}$ is about 65\% of the non-resonant coupling parameter $\varepsilon_{12}$ (see Eq.~\ref{eq:eps12ratio} and Fig.~\ref{fig:eps12res}). For the two inner planets to stay in MMR (i.e. for the resonant angle to librate) in the presence of the disruptive effect of the perturber, the mutual inclination must be sufficiently high to resist the external planet's perturbation. We have derived an analytic stability criterion Eq.~\eqref{eq:stability} when one of the inner bodies is a test mass. 

Most of our results are derived analytically, and are tested against numerical calculations based on the averaged equations of motion. Similar to the pure secular problem \citep{2017AJ....153...42L}, our results can be used to put constraints on putative cold Jupiters if a misalignment is observed between two inner planets near MMRs of order 2.

\subsection{Caveats}

Our results apply for arbitrary $I_{\rm p}$ and coupling $\varepsilon_{12, \mathrm{res}}$, provided that the mutual inclination between the inner planets is small. This is due to the approximate Hamiltonian used to model mean-motion resonance. Consequently, our results do not hold if both $I_p$ and $\varepsilon_{12, \mathrm{res}}$ are large, because in this case large misalignments can be generated. In addition, the amplitude of the mutual inclination near the secular resonance, which occurs at $\varepsilon_{12} \sim 1$ and $m_1\lesssim m_2$ (see Fig.~\ref{fig:imax}), can be large, and is therefore not constrained by our model.

More importantly, we have restricted this study to circular orbits. However, misalignment may be accompanied with non-zero eccentricities, either because they have the same origin or because they are coupled. The entanglement between eccentricity and misalignment is complex near mean-motion resonance. We will tackle this issue in a forthcoming paper.

\subsection{Application related to observations}

As noted in Section \ref{sec:intro}, a large fraction of super-Earth systems are accompanied by external cold Jupiters. By evaluating the dimensionless coupling parameters, $\varepsilon_{12}$ and $\varepsilon_\mathrm{12,res}$ (Eqs.~\ref{eq:eps12} and \ref{eq:eps12res}), one can easily assess to what extent an inclined CJ influence the mutual inclinations of the inner super-Earths. If the secular excitation by an external companion is indeed the dominant mechanism to induce misalignment in close-in planets, then this would have multiple observational applications.

On the one hand, CJ companions exterior to coplanar transiting inner planets, if exist, should be sufficiently far away and/or have sufficiently small inclination. Interestingly, \cite{2020AJ....159...38M} have found that CJ companions around multiplanet systems have statistically lower inclination than around single-transiting systems. Systems with several close-in planets are detected in transit surveys only if their mutual inclination is small. This observation is thus consistent with the coplanar super-Earth population having more aligned external companions.

\begin{figure}
	\centering
	\includegraphics[width=\linewidth]{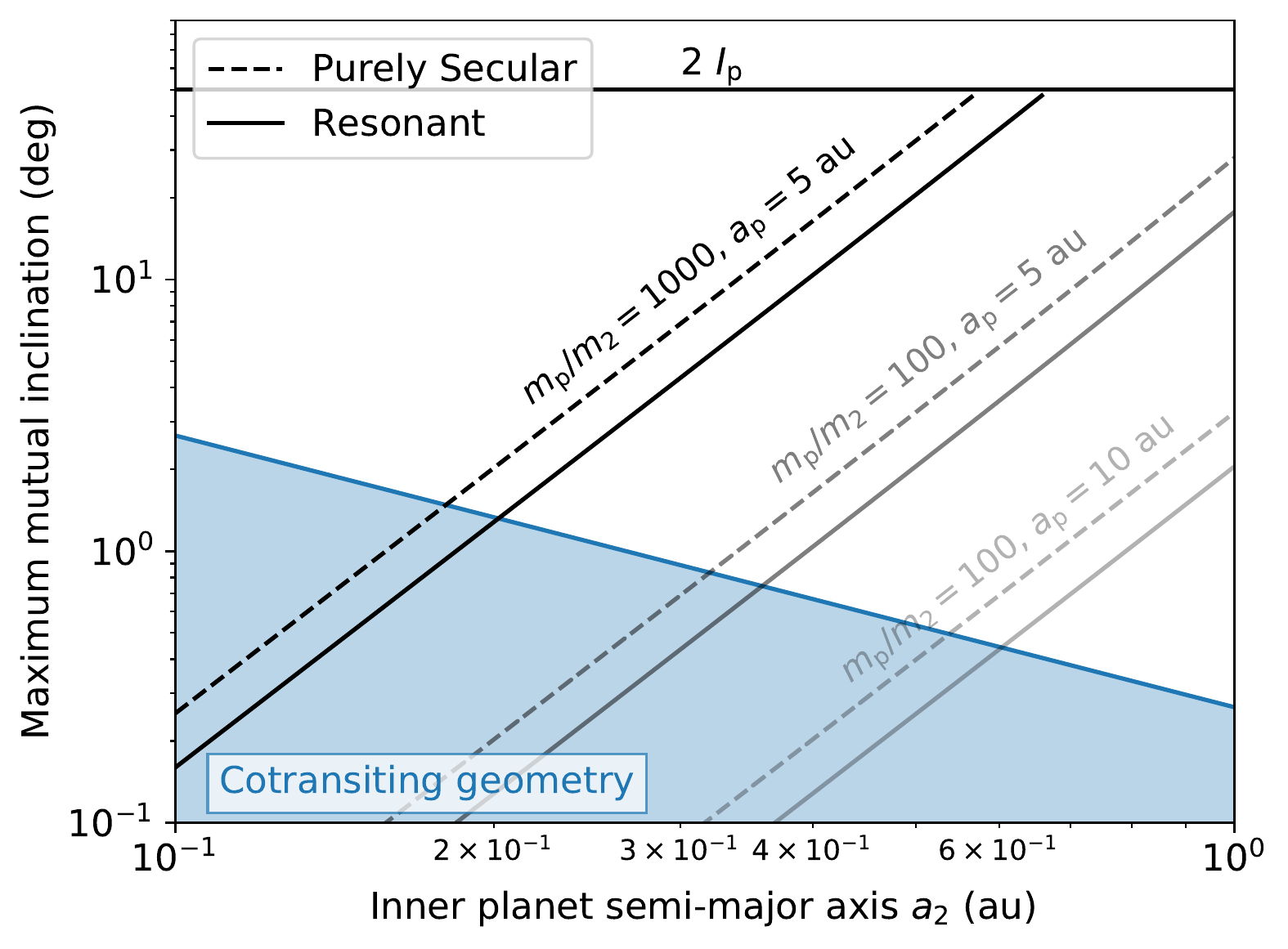}
	\caption{Maximum mutual inclination between two equal-mass planets (mass $m_1 = m_2$, and semi-major axes $a_1$, $a_2$) under the influence of a misaligned external perturber, as a function of the planet semi-major axis $a_2$. The solid and dashed lines represent the maximum inclination induced by a perturber $m_{\rm p}$ at distance $a_{\rm p}$ in the resonant (Eq.~\ref{eq:imax_comparablemass}) and non-resonant cases, respectively, with $I_{\rm p} = 25^\circ$ and $P_1/P_2 \approx 3$. The horizontal line at $2 I_{\rm p}$ is the maximum possible inclination (which is attained when the perturber is sufficiently strong). The shaded blue region represents the cotransiting zone, when the mutual inclination is below $R_\star/a_2$, for $R_\star = R_\odot$. }\label{fig:transit}
\end{figure}

On the other hand, the correlation between the existence of CJ and the mutual inclination of their inner planets can assist planet detection (see Fig.~\ref{fig:transit}). Single-transiting planets might be good candidates to look for CJ companions. In a system with several detected inner planets, measure of their mutual inclination and semi-major axis ratio (resonant or not) would give constraints on the strength of a putative perturber, that is, on its mass, semi-major axis and inclination. Conversely, knowledge of the CJ population would help us singling out systems with potentially undetected super-Earths.

\section*{Acknowledgements}

We thank the referee Alexander Mustill for a constructive report which improved the manuscript. This work has been supported in part by the NSF grant AST-17152 and NASA grant 80NSSC19K0444. The results were integrated with the \textsc{python} library \textsc{SciPy} \citep{2020NatMe..17..261V}, and the figures were made with \textsc{Matplotlib} \citep{2007CSE.....9...90H}.

\section*{Data Availability}
 
The \textsc{python} algorithm used for the integration of the differential equations characterizing the problem is available on request.



\bibliographystyle{mnras}
\bibliography{Article} 



\appendix
\onecolumn

\section{Stability around the true equilibrium}
\label{sec:stability}

In the outer test mass case (Section \ref{sec:testmass-planet}), the differential equations around the true equilibrium (Eqs.~\ref{eq:aeq_outer}--\ref{eq:Oeq_outer}) writes:
\begin{align}
\dv{\delta a}{t} &= j \omega_{\mathrm{res}} a_\mathrm{eq} I^2_{\mathrm{forced}} \delta\phi,\\
\dv{\delta I}{t} &=  \omega_{\mathrm{res}} I_{\mathrm{forced}} \delta\phi + \frac{\Delta\omega_{\rm p}}{2}\sin(2I_{\rm p}) \delta\Omega ,\\
\dv{\delta\phi}{t} &= -\frac{3}{2}n'j \left(\frac{a'}{a_\mathrm{eq}}\right)^\frac{3}{2} \frac{\delta a}{a_\mathrm{eq}} + \frac{\Delta\omega_{\rm p}}{I^2_{\mathrm{forced}}}\sin(2I_{\rm p}) \delta I,\\
\dv{\delta\Omega}{t} &= - \frac{\Delta\omega_{\rm p}}{2I^2_{\mathrm{forced}}}\sin(2I_{\rm p})\delta I,
\end{align}
\noindent which can be rewritten as
\begin{equation}
\frac{d}{dt}\left(\begin{matrix}
\delta a \\ 
\delta I\\ 
\delta\phi \\ 
\delta\Omega
\end{matrix}\right) 
= 
\left(\begin{matrix}
0 & 0 & j a_\mathrm{eq} \omega_{\mathrm{res}} I^2_{\mathrm{forced}} & 0\\
0 & 0 & \omega_{\mathrm{res}} I_{\mathrm{forced}} & {\Delta\omega_{\rm p}}\sin(2I_{\rm p})/2\\ 
\frac{3}{2}n'j {a^{'\frac{3}{2}}}/{a_\mathrm{eq}^\frac{5}{2}} & {\Delta\omega_{\rm p}}\sin(2 I_{\rm p})/{I^2_{\mathrm{forced}}} & 0 & 0\\
0 & - {\Delta\omega_{\rm p}}\sin(2I_{\rm p})/{2I^2_{\mathrm{forced}}} & 0 & 0
\end{matrix}\right)
\left(\begin{matrix}
\delta a \\ 
\delta I \\ 
\delta\phi \\ 
\delta\Omega
\end{matrix}\right) ,
\end{equation}
The eigenvalues have the form:
\begin{equation}
\lambda_\pm = \pm \frac{1}{2\sqrt{2}} \sqrt{x \pm \sqrt{x^2 - y}},
\end{equation}
where, assuming $(a'/a_{\mathrm{eq}})^\frac{3}{2} \approx (j-2)/j$:
\begin{align}
x &= 8 \omega_{\mathrm{res}} \omega_{\mathrm{prec}} - 4  \omega^2_{\mathrm{prec}} - \frac{3}{2}n'j(j-2)\Delta\omega_{\rm p}^2\sin^2(2I_{\rm p})\frac{\omega_{\mathrm{res}}}{\omega_{\mathrm{prec}}^2},\\
y &= 24 j (j-2) \Delta\omega_{\rm p}^2\sin^2(2I_{\rm p}) \omega_{\mathrm{res}}.
\end{align}
As we always have (for whatever $j$) $\omega_{\mathrm{prec}} > 2\omega_{\mathrm{res}}$, then $x<0$. The stability of the equilibrium then depends on $x^2 - y$: unstable when negative, stable when positive. 
\begin{align}
&x^2 - y < 0 \quad\text{(instability)}\\
&\iff \frac{3}{2}n'j(j-2)\frac{\omega_{\mathrm{res}}}{\omega_{\mathrm{prec}}^2}\Delta\omega_{\rm p}^2\sin^2(2I_{\rm p})- \sqrt{24 n' j (j-2)\omega_{\mathrm{res}}}\Delta\omega_{\rm p}\sin(2I_{\rm p}) -8 \omega_{\mathrm{res}} \omega_{\mathrm{prec}} + 4  \omega^2_{\mathrm{prec}} < 0\label{eq:instability}.
\end{align}
This expression is a polynomial of degree $2$ in $\Delta\omega_{\rm p}\sin(2I_{\rm p})$, so that we can derive an instability condition:
\begin{equation}
\left(\frac{\omega_{\mathrm{prec}}}{n'}\right)^\frac{3}{2}\left(\sqrt{\frac{\omega_{\mathrm{prec}}}{2\omega_{\mathrm{res}}}}-1\right)\sqrt{\frac{16}{3j(j-2)}} < \frac{\Delta\omega_{\rm p}\sin(2I_{\rm p})}{n'}< \left(\frac{\omega_{\mathrm{prec}}}{n'}\right)^\frac{3}{2}\left(\sqrt{\frac{\omega_{\mathrm{prec}}}{2\omega_{\mathrm{res}}}}+1\right)\sqrt{\frac{16}{3j(j-2)}}.
\end{equation}
If $\varepsilon_\mathrm{12,res}$ is small, it reduces to:
\begin{equation}
\sqrt\frac{\omega_{\mathrm{prec}}}{n'}\left(\sqrt{\frac{\omega_{\mathrm{prec}}}{2\omega_{\mathrm{res}}}}-1\right)\sqrt{\frac{16}{3j(j-2)}} < \varepsilon_\mathrm{12,res}\sin(2I_{\rm p})< \sqrt\frac{\omega_{\mathrm{prec}}}{n'} \left(\sqrt{\frac{\omega_{\mathrm{prec}}}{2\omega_{\mathrm{res}}}}+1\right)\sqrt{\frac{16}{3j(j-2)}} \approx \sqrt{\mu'}.\label{eq:instability_strongcoupling}
\end{equation}
An example of a stable evolution with $\varepsilon_\mathrm{12,res}\sin(2I_{\rm p})$ below the lower limit is shown in Fig.~\ref{fig:example_stabilityisland}. It can be shown by studying numerically the roots of Eq.~\eqref{eq:instability} that, for reasonable mass ratios $\mu' \lesssim 10^{-3}$, the stability is ensured for any coupling if $\varepsilon_\mathrm{12,res}$ and $I_{\rm p}$ roughly satisfy:
 \begin{equation}
 \varepsilon_\mathrm{12,res}\sin(2I_{\rm p})> \sqrt{\mu'}.
 \end{equation}

The same reasoning can be held for the inner test mass case, leading to the same instability condition.

\section{Rotating Frame}
\label{sec:rotatingframe}

The evolution of the total angular momentum $\vec L = \vec L_1 + \vec L_2 = L \vec{\hat l}$ is governed by
\begin{align}
\dv{\vec{L}}{t} &= (\omega_{1{\rm p}}\cos\theta_{1{\rm p}}\vec{L_1} + \omega_{2{\rm p}}\cos\theta_{2{\rm p}}\vec{L_2})\wedge \vec{\hat l_{\rm p}}.
\end{align}
If the relative inclination $I$ between $\vec{L_1}$ and $\vec{L_2}$ is small, then:
\begin{align}
\vec{L_1} &= L_1 \vec{\hat l} + (\vec{L_1}-\frac{L_1}{L}\vec{L}) = L_1 \vec{\hat l} + \frac{L_2\vec{L_1}-L_1\vec{L_2}}{L} + O(I^2) = L_1 \vec{\hat l} +  O(I^2).
\end{align}
Moreover, $\cos\theta_{1{\rm p}} = \cos\theta_{2{\rm p}} + O(I) = \cos I_{\rm p} + O(I)$, where $I_{\rm p}$ is the angle between $\vec L$ and $\vec L_{\rm p}$. Thus
\begin{align}
\dv{\vec{L}}{t} &= \frac{\omega_{1{\rm p}} L_1 + \omega_{2{\rm p}} L_2}{L}\cos I_{\rm p} \vec{L}\times \vec{\hat l_{\rm p}}.
\end{align}
Let us now consider the possible variation of $I_{\rm p}$. We have:
\begin{align}
L \cos I_{\rm p} = L_1 \cos\theta_{1{\rm p}} + L_2\cos\theta_{2{\rm p}}.
\end{align}
The relative inclinations $\theta_{1{\rm p}}$ and $\theta_{2{\rm p}}$ are unchanged by the perturber, but vary as $m_1$ and $m_2$ interact. However, we have:
\begin{align}
\dv{}{t} L_1 \cos\theta_{1{\rm p}} = -\dv{}{t} L_2\cos\theta_{2{\rm p}},
\end{align}
so that
\begin{align}
\dv{L \cos I_{\rm p}}{t} = 0.
\end{align}
After some algebra, using $L^2 = L_1^2 + L_2^2 + 2 L_1 L_2 \cos I$:
\begin{align}
\dv{I_{\rm p}}{t} &= -\frac{L_1 L_2}{L} \frac{\sin I}{\tan I_{\rm p}} \dv{I}{t}\nonumber,\\
&= -\frac{L_1 L_2}{L} \sin I \cos(\Omega_2-\Omega_1) \cos I_{\rm p} (\omega_{2{\rm p}}\cos\theta_{2{\rm p}}-\omega_{1{\rm p}}\cos\theta_{1{\rm p}}).
\end{align}
Inspiring from Section \ref{sec:testmass-planet}, we suppose that $\cos I_{\rm p}$ stays roughly constant and $\sin I \cos(\Omega_2-\Omega_1) = I_\mathrm{free} \sin(\omega_{\mathrm{prec}}t + \psi)$, and we get:
\begin{align}
\Delta I_{\rm p} \approx \frac{L_1 L_2}{L} \varepsilon_{12,\mathrm{res}}  I_\mathrm{free}\cos^2 I_{\rm p}.
\end{align}
All in all, we can suppose $I_{\rm p}$ constant only if $L_1$ or $L_2$ is zero (test mass), or if $\varepsilon_{12}$ is small (strong coupling).

\section{Full set of equations for the comparable mass case}
\label{sec:appendixcomparablemass}

\subsection{General coupling, small $I_{\rm p}$}

From the Hamilton-Jacobi equations applied to the Hamiltonian Eq.~\eqref{eq:hamiltonian}, we get the evolution of the angles:
\begin{align}
\dv{\lambda_1}{t} &= \pdv{H}{L_1}\big|_{Z_1} = \pdv{H}{L_1}\big|_{I_1} - \frac{I_1}{2Z_1} \pdv{H}{I_1}\big|_{L_1} = n_1 + O(\mu', \mu_{\rm p}),\\
\dv{\lambda_2}{t} &= \pdv{H}{L_2}\big|_{Z_2} = \pdv{H}{L_2}\big|_{I_2} - \frac{I_2}{2Z_2} \pdv{H}{I_2}\big|_{L_2} = n_2 + O(\mu', \mu_{\rm p}),\\
\dv{\Omega_1}{t} &= -\pdv{H}{Z_1}\big|_{L_1} = -\frac{1}{L_1 I_1} \pdv{H}{I_1}\big|_{L_1}\nonumber\\
&= - \omega_{12} + \omega_{12} \frac{I_2}{I_1}\cos(\Omega_2-\Omega_1) + \omega_{\mathrm{12,res}}  \cos\phi_1 - \omega_{\mathrm{12,res}} \frac{I_2}{I_1}\cos\left(\frac{\phi_1+\phi_2}{2}\right)\nonumber\\
& + \frac{\omega_{1{\rm p}}}{I_1} \cos\theta_{1{\rm p}} \left(\cos(\Omega_1-\Omega_{\rm p})\sin I_{\rm p} - I_1 \cos I_{\rm p}\right),\label{eq:dO1dt}\\
\dv{\Omega_2}{t} &= -\pdv{H}{Z_2}\big|_{L_2} = -\frac{1}{L_2 I_2} \pdv{H}{I_2}\big|_{L_2}\nonumber\\
&= - \omega_{21} + \omega_{21} \frac{I_1}{I_2}\cos(\Omega_2-\Omega_1) + \omega_{\mathrm{21,res}}  \cos\phi_2 - \omega_{\mathrm{21,res}} \frac{I_1}{I_2}\cos\left(\frac{\phi_1+\phi_2}{2}\right)\nonumber\\
& + \frac{\omega_{2{\rm p}}}{I_2} \cos\theta_{2{\rm p}} \left(\cos(\Omega_2-\Omega_{\rm p})\sin I_{\rm p} - I_2 \cos I_{\rm p}\right).\label{eq:dO2dt}
\end{align}
and the evolution of the momenta
\begin{align}
\dv{L_1}{t} &= -\pdv{H}{\lambda_1} = \frac{j-2}{2} \omega_{12,\mathrm{res}}L\left(I_1^2 \sin\phi_1 + I_2^2\sin\phi_2 - 2 I_1 I_2 \sin\left(\frac{\phi_1+\phi_2}{2}\right)\right),\\
\dv{L_2}{t} &= -\pdv{H}{\lambda_2} = -\frac{j}{2} \omega_{21,\mathrm{res}}L\left(I_1^2 \sin\phi_1 + I_2^2\sin\phi_2 - 2 I_1 I_2 \sin\left(\frac{\phi_1+\phi_2}{2}\right)\right),\\
\dv{Z_1}{t} &= \pdv{H}{\Omega_1} = -\omega_{12}L_1 I_1 I_2 \sin(\Omega_2-\Omega_1) - \omega_{12,\mathrm{res}}L\left(I_1^2 \sin\phi_1 - I_1 I_2 \sin\left(\frac{\phi_1+\phi_2}{2}\right)\right) + \omega_{1{\rm p}}\cos\theta_{1{\rm p}} L_1 I_1 \sin I_{\rm p} \sin\Omega_1,\\
\dv{Z_2}{t} &= \pdv{H}{\Omega_2} = \omega_{21}L_2 I_1 I_2 \sin(\Omega_2-\Omega_1) - \omega_{21,\mathrm{res}}L\left(I_2^2 \sin\phi_2 - I_1 I_2 \sin\left(\frac{\phi_1+\phi_2}{2}\right)\right) + \omega_{2{\rm p}}\cos\theta_{2{\rm p}} L_2 I_2 \sin I_{\rm p} \sin\Omega_2.
\end{align}
Deriving the semi-major axes and inclinations from the momenta, we finally get
\begin{align}
\dv{a_1}{t} &= \frac{2 L_1}{m_1^2 GM} \dv{L_1}{t}\nonumber\\
&= (j-2)\omega_{12,\mathrm{res}}  a_1 \left(I_1^2 \sin\phi_1 + I_2^2\sin\phi_2 - 2 I_1 I_2 \sin\left(\frac{\phi_1+\phi_2}{2}\right)\right),\\
\dv{a_2}{t} &= \frac{2 L_2}{m_2^2 GM} \dv{L_2}{t}\nonumber\\
&= -j  \omega_{\mathrm{21,res}} a_2\left(I_1^2 \sin\phi_1 + I_2^2\sin\phi_2 - 2 I_1 I_2 \sin\left(\frac{\phi_1+\phi_2}{2}\right)\right),\\
\dv{I_1}{t} &= \frac{1}{L_1 I_1} \dv{Z_1}{t} - \frac{I_1}{2L_1} \dv{L_1}{t}\nonumber\\
&=  -\omega_{12} I_2 \sin(\Omega_2-\Omega_1) - \omega_{\mathrm{12,res}}\left(I_1 \sin\phi_1 - I_2 \sin\left(\frac{\phi_1+\phi_2}{2}\right)\right) + \omega_{1{\rm p}}\cos\theta_{1{\rm p}} \sin I_{\rm p} \sin(\Omega_1-\Omega_{\rm p}),\label{eq:dI1dt}\\
\dv{I_2}{t} &= \frac{1}{L_2 I_2} \dv{Z_2}{t} - \frac{I_2}{2L_2} \dv{L_2}{t}\nonumber\\
&=  \omega_{21} I_1 \sin(\Omega_2-\Omega_1) - \omega_{\mathrm{21,res}}\left(I_2 \sin\phi_2 - I_1 \sin\left(\frac{\phi_1+\phi_2}{2}\right)\right) + \omega_{2{\rm p}}\cos\theta_{2{\rm p}} \sin I_{\rm p} \sin(\Omega_2-\Omega_{\rm p}).\label{eq:dI2dt}
\end{align}
From there, we can get to six equations by replacing the evolution of $a_1$ and $a_2$ by $\alpha = a_1/a_2$ and the evolution of $\lambda_1$ and $\lambda_2$ by $j\lambda_2 - (j-2)\lambda_1$:
\begin{align}
&\frac{1}{\alpha}\dv{\alpha}{t} = \frac{1}{a_1}\dv{a_1}{t} - \frac{1}{a_2}\dv{a_2}{t} = \left((j-2) \omega_{\mathrm{12,res}} + j \omega_{\mathrm{21,res}}\right) \left(I_1^2 \sin\phi_1 + I_2^2\sin\phi_2 - 2 I_1 I_2 \sin\left(\frac{\phi_1+\phi_2}{2}\right)\right)\label{eq:dalphadt},\\
&\dv{}{t}\left[j\lambda_2 - (j-2)\lambda_1 \right] = jn_2 - (j-2) n_1 = -\Delta n.\label{eq:dthetadt}
\end{align}
The final set of equations is Eqs.~\eqref{eq:dalphadt}, \eqref{eq:dthetadt}, \eqref{eq:dI1dt}, \eqref{eq:dI2dt}, \eqref{eq:dO1dt}, \eqref{eq:dO2dt}.

\subsection{Strong coupling}

Assuming $\vec L$ is rotating around $\vec L_{\rm p}$ at the rate $-\omega_{\rm p} \cos I_{\rm p}$ with $I_{\rm p}$ constant, we can reduce the problem to four equations by choosing as the reference plane the plane perpendicular to $\vec{L}$. New terms have then to be added to take into account the rotation:
\begin{align}
\dv{\Omega_1}{t}_\mathrm{new} &= \dv{\Omega_1}{t} - \frac{\omega_{\rm p}}{I_1} \cos I_{\rm p} \left(\cos(\Omega_1-\Omega_{\rm p}) \sin I_{\rm p} - I_1 \cos I_{\rm p}\right),\\
\dv{\Omega_2}{t}_\mathrm{new} &= \dv{\Omega_2}{t} - \frac{\omega_{{\rm p}}}{I_2} \cos I_{\rm p} \left(\cos(\Omega_2-\Omega_{\rm p}) \sin I_{\rm p} - I_2 \cos I_{\rm p}\right),\\
\dv{I_1}{t}_\mathrm{new} &= \dv{I_1}{t} - \omega_{{\rm p}}\cos I_{\rm p} \sin I_{\rm p} \sin(\Omega_1-\Omega_{\rm p}),\\
\dv{I_2}{t}_\mathrm{new} &= \dv{I_2}{t} - \omega_{{\rm p}}\cos I_{\rm p} \sin I_{\rm p} \sin(\Omega_2-\Omega_{\rm p}).
\end{align}
Moreover, our choice of frame ensures that $I = I_1+I_2$, $L_1\sin I_1 = L_2\sin I_2$ and $\Omega_1 = \Omega_2 + \upi$. We can thus make use of the following relations:
\begin{align}
&\dv{I}{t} = \dv{I_1}{1} + \dv{I_2}{t},\\
&\omega_{{\rm p}}-\omega_{1{\rm p}} = \frac{L_2}{L}\Delta\omega_{\rm p},\\
&\omega_{2{\rm p}}-\omega_{\rm p} = \frac{L_1}{L}\Delta\omega_{\rm p},\\
&\frac{\omega_{2{\rm p}}-\omega_{\rm p}}{I_2} = \frac{\omega_{{\rm p}}-\omega_{1{\rm p}}}{I_1} =\frac{\Delta\omega_{\rm p}}{I}.
\end{align}
Finally, we get:
\begin{align}
\dv{\alpha}{t} &= \alpha \left((j-2) \omega_{\mathrm{12,res}} + j \omega_{\mathrm{21,res}}\right) I^2 \sin\phi,\\
\dv{I}{t} &= - (\omega_{\mathrm{12,res,1}}+\omega_{\mathrm{21,res}}) I \sin\phi -  \frac{\Delta\omega_{\rm p}}{2}\sin(2I_{\rm p})\sin(\Omega_1-\Omega_{\rm p}),\\
\dv\phi{t} &= -\Delta n - 2\left(\omega_{\mathrm{12,res}} + \omega_{\mathrm{21,res}}\right) \cos\phi + \frac{\Delta\omega_{\rm p}}{I}\cos(\Omega_1-\Omega_{\rm p})\sin(2I_{\rm p}),\\
\dv{\Omega_1}{t} &= - (\omega_{12}+\omega_{21}) +  \left(\omega_{\mathrm{12,res}} + \omega_{\mathrm{21,res}}\right) \cos\phi+\frac{L_2}{L}\Delta\omega_{\rm p} \cos^2 I_{\rm p} - \frac{\Delta\omega_{\rm p}}{2I} \cos(\Omega_1-\Omega_{\rm p})\sin (2I_{\rm p}).
\end{align}
In order to ensure that $\dv{\Omega_1}{t} = \dv{\Omega_2}{t}$, $\Delta\omega_{\rm p}$ should be negligible with respect to $\omega_{12}+\omega_{21}+\omega_{12, \mathrm{res}}+\omega_{21, \mathrm{res}}$ (strong coupling limit). 


\bsp	
\label{lastpage}
\end{document}